\documentclass[12pt,a4paper]{article}
\usepackage{amsmath,amssymb}
\usepackage{slashed}
\usepackage[dvipdfmx]{graphicx}
\usepackage[dvips]{color}
\begin{document}
\newcounter{Series}
\setcounter{Series}{2}
\title{
A Diagrammer's Note on 
Superconducting Fluctuation Transport 
for Beginners: \\
\Roman{Series}. Hall and Nernst Effects \\
with Perturbational Treatment \\
of Magnetic Field }

\author{
O. Narikiyo
\footnote{
Department of Physics, 
Kyushu University, 
Fukuoka 819-0395, 
Japan}
}

\date{
(Sep. 2, 2026)
}
\maketitle
\begin{abstract}
A diagrammatic approach 
based on thermal Green function 
to superconducting fluctuation transport 
is reviewed focusing on Hall and Nernst effects. 
The treatment of weak magnetic field is carefully discussed 
within the linear order perturbation. 

In the Appendix 
the linear response theory for the DC Hall conductivity 
of the Dirac fermion in 2+1 space-time dimensions is reviewed. 
One focus is the Chern-Simons effective action for the gauge field. 
Another is the exact formula by Ishikawa and Matsuyama. 
\end{abstract}
\vskip 20pt

\section{Introduction}

This Note is the second part of the series\footnote{ 
In this Note I shall quote the first Note as [I]. 

[I] $\equiv$ Narikiyo: arXiv:1112.1513. } 
and I expect that you have already read the first one [I]. 
In this second Note I shall discuss the effect of the magnetic field 
perturbationally in the weak field limit. 
The finite magnetic field shall be discussed 
non-perturbationally in the third Note. 
The symbols that have appeared in [I] 
are used here without explanations. 
Since the introduction to the series has been given in [I], 
let us start at once. 

\newpage

\section{Boltzmann Transport: Relaxation-Time Approximation}

In the relaxation-time approximation 
of the Boltzmann transport\footnote{ 
The Boltzmann transport is formulated in terms of ${\bf E}$ and ${\bf H}$ 
so that it is {\bf gauge-invariant}. See footnote 12 in [I]. } 
the expectation value of the charge current ${\bf J}^e$ 
and the heat current ${\bf J}^Q$ 
in the linear order of the uniform electric field ${\bf E}$ 
and the uniform magnetic field ${\bf H}$ are given as\footnote{ 
See (44) and (45) in [I]. } 
\begin{equation}
{\bf J}^e = - 2 { e^3 \over m } \sum_{\bf p} v_x^2 \tau^2 
\Big( - { \partial f(\xi_{\bf p}) \over \partial \xi_{\bf p} } \Big) 
\big( {\bf H}\times{\bf E} \big), 
\label{J^e(H,E)} 
\end{equation}
and 
\begin{equation}
{\bf J}^Q = - 2 { e^2 \over m } \sum_{\bf p} v_x^2 \tau^2 \xi_{\bf p} 
\Big( - { \partial f(\xi_{\bf p}) \over \partial \xi_{\bf p} } \Big) 
\big( {\bf H}\times{\bf E} \big), 
\label{J^Q(H,E)} 
\end{equation}
where both ${\bf E}$ and ${\bf H}$ are static. 
These describe the currents in Hall and Nernst effects. 
Although formally exact results have been obtained,\footnote{ 
The exact formula for ${\bf J}^e$ via the Boltzmann equation 
is given by Kotliar, Sengupta and Varma: Phys. Rev. B {\bf 53}, 3573 (1996) 
as has been discussed in \S 6 of [I] 
and it is also derived from the Fermi-liquid theory in [KY]. 

The exact formula for ${\bf J}^Q$ via the Boltzmann equation 
is given by Pikulin, Hou and Beenakker: Phys. Rev. B {\bf 84}, 035133 (2011) 
and it is also derived from the Fermi-liquid theory in [Kon]. 

In the Fermi-liquid theory 
the effect of scatterings beyond the relaxation-time approximation 
is taken into account as the vertex correction. 

[KY] $\equiv$ Kohno and Yamada: Prog. Theor. Phys. {\bf 80}, 623 (1988). 

[Kon] $\equiv$ Kontani: Phys. Rev. B {\bf 67}, 014408 (2003). } 
I shall only discuss the results within the relaxation-time approximation 
in the following sections. 

\color{red}
Before ending this section, I shall add some comments on the footnote 4. 
The charge current is given by [I]-(37) and the heat current is given by [I]-(38). 
In these two equations the common unknown quantity 
is the deviation of the distribution $g_{\bf p}$. 
In the presence of the magnetic field $g_{\bf p}$ is given by [I]-(52). 
As shown in [I]-(43), 
$e {\bf E}$ in [I]-(52) is replaced by $e {\bf E} - \xi_{\bf p} \nabla T / T$ 
in the presence of the temperature gradient. 
These are the results of the $A$-matrix formula by Kotliar, Sengupta and Varma. 
On the other hand, Pikulin, Hou and Beenakker exploited the vector mean free path. 
I have discussed such a treatment in arXiv:1607.04705v2. 
\color{black}

\newpage 
\section{Quasi-particle Transport: Relaxation-Time Approximation}

We consider the linear response to electric field 
in the presence of magnetic field. 
The effect of the magnetic field is treated perturbationally 
in the weak-field limit. 

The expectation value of the charge current ${\bf J}^e$ 
is expressed as the linear response to electric field 
\begin{equation}
J^e_\mu({\bf k}, \omega) = 
\sum_\nu \sigma_{\mu\nu}({\bf k}, \omega) E_\nu({\bf k} \!=\! 0, \omega), 
\end{equation}
where the electric field ${\bf E}$ is uniform\footnote{ 
$ E_\nu({\bf k} \!=\! 0, \omega) $ should be used in (68) and (78) of [I]. }  
and the ${\bf k}$-dependence\footnote{ 
We do not have to consider the ${\bf k}$-dependence in \S 8 of [I] 
where the magnetic field is absent. } 
comes from 
the vector potential ${\bf A}({\bf x})$ introduced as 
\begin{equation}
{\bf A}({\bf x}) = {\bf A} e^{i{\bf k}\cdot{\bf x}}, 
\end{equation}
with a constant vector ${\bf A}$. 
The magnetic field ${\bf H}$ in the limit of ${\bf k} \rightarrow 0$ 
is expressed as 
\begin{equation}
{\bf H} = \nabla \times {\bf A}({\bf x}) = i ({\bf k} \times {\bf A}). 
\label{H=rotA} 
\end{equation}
The conductivity tensor is given 
by the Kubo formula\footnote{
I follow the derivation of the Hall conductivity by [FEW]. 
Their derivation for the case of nearly free electrons are extended 
to the case of interacting electrons by [KY] 
on the basis of the Fermi-liquid theory. 
These works are formulated in fully {\bf gauge-invariant} manner. 

[FEW] $\equiv$ 
Fukuyama, Ebisawa and Wada: Prog. Theor. Phys. {\bf 42}, 494 (1969). } 
(linear response theory) 
\begin{equation}
\sigma_{\mu\nu}({\bf k}, \omega) = { 1 \over i\omega } 
\big[ \Phi_{\mu\nu}^e({\bf k}, \omega + i\delta) 
- \Phi_{\mu\nu}^e({\bf k}, i\delta) \big], 
\end{equation}
where 
\begin{equation}
\Phi_{\mu\nu}^e({\bf k}, i\omega_\lambda) = 
\int_0^\beta d \tau \, e^{i\omega_\lambda\tau} 
\big\langle T_\tau \big\{ \, 
j^H_\mu({\bf k}; \tau) \, j^H_\nu({\bf k} \!=\! 0) \big\} \big\rangle. 
\label{def-Phi-e} 
\end{equation}
Here ${\bf J}^H$ is the charge current in the presence of the magnetic field 
and its Fourier component is given by\footnote{ 
Here the diamagnetic current density is transformed by the integral 
\begin{equation}
- {e^2 \over m} \sum_\sigma \int d^3 x \, e^{-i{\bf q}\cdot{\bf x}} 
\psi_\sigma^\dag({\bf x}) \psi_\sigma({\bf x}) {\bf A}({\bf x}), 
\nonumber 
\end{equation}
with 
\begin{equation}
\rho({\bf q}) 
= e \sum_\sigma \int d^3 x \, e^{-i{\bf q}\cdot{\bf x}} 
\psi_\sigma^\dag({\bf x}) \psi_\sigma({\bf x}). 
\nonumber 
\end{equation}
} 
\begin{equation}
j^H_\mu({\bf q}) 
= j^e_\mu({\bf q}) - {e \over m} \rho({\bf q}-{\bf k}) A_\mu, 
\end{equation}
where\footnote{
It should be noted that each term in ${\bf j}^e({\bf q})$ 
is proportional to the summation of incoming and outgoing velocities 
${\bf v}_{\bf p} + {\bf v}_{\bf p+q}$. } 
\begin{align}
{\bf j}^e({\bf q}) 
& = e \sum_\sigma \sum_{\bf p} { 1 \over 2 } 
\big( {\bf v}_{\bf p} + {\bf v}_{\bf p+q} \big) 
c_{{\bf p}\sigma}^\dag c_{{\bf p+q}\sigma} 
\nonumber \\
& = {e \over m} \sum_\sigma \sum_{\bf p} 
\Big( {\bf p} + { {\bf q} \over 2 }\Big) 
c_{{\bf p}\sigma}^\dag c_{{\bf p+q}\sigma}, 
\label{j^e(q)} 
\end{align}
and 
\begin{equation}
\rho({\bf q}) = e \sum_\sigma \sum_{\bf p} 
c_{{\bf p}\sigma}^\dag c_{{\bf p+q}\sigma}. 
\end{equation}
In this note $e$ is chosen to be negative $e<0$. 
It should be noted that in (\ref{def-Phi-e}) 
the ${\bf k}$-independence of $j^H_\nu({\bf k} \!=\! 0)$ 
results from the coupling to the uniform electric field\footnote{ 
The uniform electric field ${\bf E}$ is expressed 
by the uniform vector potential ${\bf A}_0$ as ${\bf E}=i\omega{\bf A}_0$ 
where we have put $\phi=0$ with $\phi$ being the scalar potential. 
The current ${\bf j}^H({\bf k} \!=\! 0)$ couples to ${\bf A}_0$. } 
and the ${\bf k}$-dependence of $j^H_\mu({\bf k}; \tau)$ represents 
the magnetic-field dependence of the observed current. 

The (imaginary) time-dependence is introduced by 
\begin{equation}
j^H_\mu({\bf k}; \tau) = e^{K\tau} \, j^H_\mu({\bf k}) \,  e^{-K\tau}, 
\end{equation}
where the coupling to the magnetic field\footnote{ 
It is sufficient to consider the coupling 
\begin{equation}
- \int d^3x \ {\bf j}^e({\bf x}) \cdot {\bf A}({\bf x}), 
\nonumber 
\end{equation}
for the discussion of the observed current linear in ${\bf A}$. }
\begin{equation}
- {\bf j}^e( - {\bf k}) \cdot {\bf A}, 
\label{jA-int} 
\end{equation}
is added to $K$ in (1) of [I]. 
By the coupling to the vector potential 
the momentum of the electron increases by ${\bf k}$ 
as seen from (\ref{j^e(q)}). 
Thus the coupling leads to the electron propagator 
off-diagonal in the momentum. 

In the following we focus on the case of nearly free electrons 
where the electron propagator in the absence of magnetic field is given by 
\begin{equation}
G({\bf p}, i\varepsilon_n) = 
{ 1 \over i{\tilde \varepsilon_n} - \xi_{\bf p} }, 
\label{thermal-G} 
\end{equation}
with 
\begin{equation}
{\tilde \varepsilon_n} \equiv 
\varepsilon_n + {1 \over 2\tau} {\rm sgn}(\varepsilon_n). 
\end{equation}
Here $\tau$ is the life-time due to quasi-elastic scatterings. 

\vskip 8mm
\begin{figure}[htbp]
\begin{center}
\includegraphics[width=13.8cm]{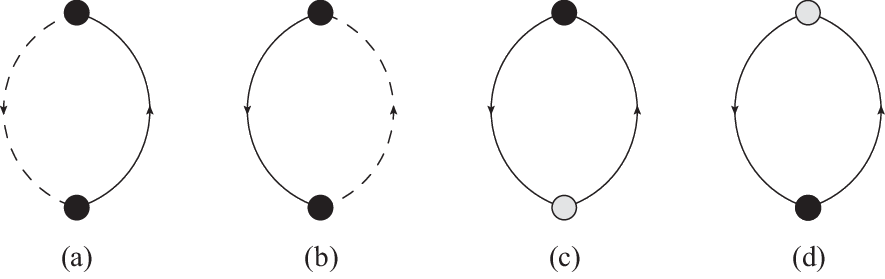}
\caption{Diagrams with ${\bf A}$-linear contributions: 
(a) The solid line in upward direction represents 
the electron propagator $-G({\bf p+k/2}, i\varepsilon_n+i\omega_\lambda)$ 
and the broken line in downward direction 
represents $-G({\bf p+k/2}\leftarrow{\bf p-k/2}, i\varepsilon_n)$. 
Throughout the series of three Notes the Zeeman splitting is neglected 
so that the propagators for up-spin electrons and for down-spin electrons 
are degenerate. 
The upper black circle represents a component of $j^e_\mu({\bf k})$ 
where the momentum of the electron changes 
from ${\bf p+k/2}$ to ${\bf p-k/2}$ 
and the lower black circle represents a component of $j^e_\nu(0)$ 
where the momentum does not change. 
(b) The broken line in upward direction is 
$-G({\bf p+k/2}\leftarrow{\bf p-k/2}, i\varepsilon_n+i\omega_\lambda)$ 
and the solid line in downward direction 
is $-G({\bf p-k/2}, i\varepsilon_n)$. 
The upper black circle is $j^e_\mu({\bf k})$ 
and the lower black circle is $j^e_\nu(0)$. 
(c) The solid line in upward direction is 
$-G({\bf p+k/2}, i\varepsilon_n+i\omega_\lambda)$ 
and the solid line in downward direction is $-G({\bf p-k/2}, i\varepsilon_n)$. 
The upper black circle is $j^e_\mu({\bf k})$ 
and the lower gray circle is $ - (e / m) \rho(-{\bf k}) A_\nu$ 
where the momentum changes from ${\bf p-k/2}$ to ${\bf p+k/2}$. 
(d) The solid line in upward direction is 
$-G({\bf p}, i\varepsilon_n+i\omega_\lambda)$ 
and the solid line in downward direction is $-G({\bf p}, i\varepsilon_n)$. 
The upper gray circle is $ - (e / m) \rho(0) A_\mu$, 
where the momentum does not change, 
and the lower gray circle is $j^e_\nu(0)$. 
The integral of this diagram in terms of ${\bf p}$ 
is odd under the variable change $p_\nu \rightarrow - p_\nu$ 
and vanishes so that (d) does not contribute to the conductivity. } 
\label{fig:4-Loop}
\end{center}
\end{figure}

The Feynman diagrams for the current-current correlation function\footnote{
The procedure described in the footnote 20 of [I] 
can be repeated in terms of 
the off-diagonal propagator (\ref{off-diagonal-prop}). 
Here we put 
${\mathbb F}(\tau,\tau) \equiv 
\langle {\bf j}^e({\bf k}; \tau) {\bf j}^e({\bf k} \!=\! 0) \rangle$. 
Neglecting the vertex correction 
\begin{equation}
{\mathbb F}(\tau,\tau) = 2e \sum_{\bf p} \sum_{\bf p'} 
{{\bf v}_{\bf p+k/2}+{\bf v}_{\bf p-k/2} \over 2 } 
\big\langle a_{\bf p-k/2}^\dag(\tau) a_{\bf p+k/2}(\tau) 
a_{\bf p'}^\dag(0) a_{\bf p'}(0) \big\rangle {\bf v}_{\bf p'} 
\nonumber 
\end{equation}
is factorized as 
\begin{align}
{\mathbb F}(\tau,\tau) = 2e \sum_{\bf p} { {\bf p} \over m } \Big[ 
& \big\langle a_{\bf p+k/2}(\tau) a_{\bf p+k/2}^\dag(0) \big\rangle 
\big\langle a_{\bf p-k/2}^\dag(\tau) a_{\bf p+k/2}(0) \big\rangle 
{\bf v}_{\bf p+k/2} 
\nonumber \\
+ & \big\langle a_{\bf p+k/2}(\tau) a_{\bf p-k/2}^\dag(0) \big\rangle 
\big\langle a_{\bf p-k/2}^\dag(\tau) a_{\bf p-k/2}(0) \big\rangle 
{\bf v}_{\bf p-k/2} \Big] 
\nonumber
\end{align}
within the linear order of ${\bf A}$ (See (\ref{GAG})). 
Since 
\begin{equation}
\big\langle a_{\bf p}(\tau) a_{\bf p'}^\dag(0) \big\rangle 
= - G({\bf p}\leftarrow{\bf p'},\tau), 
\ \ \ \ \ \ \ \ \ \ 
\big\langle a_{\bf p}^\dag(\tau) a_{\bf p}(0) \big\rangle 
= G({\bf p}, - \tau), 
\nonumber 
\end{equation}
for $\tau > 0$, 
\begin{align}
{\mathbb F}(\tau,\tau) = - 2e \sum_{\bf p} { {\bf p} \over m } \Big[ 
& G({\bf p+k/2},\tau) G({\bf p+k/2}\leftarrow{\bf p-k/2}, - \tau) 
{ {\bf p+k/2} \over m } 
\nonumber \\ + 
& G({\bf p+k/2}\leftarrow{\bf p-k/2},\tau) G({\bf p-k/2}, - \tau) 
{ {\bf p-k/2} \over m } \Big]. 
\nonumber 
\end{align}
Introducing the Fourier transforms 
\begin{equation}
{\mathbb F}(\tau,\tau) = {1 \over \beta} \sum_{m} 
{\mathbb F}(i\omega_m) e^{-i\omega_m \tau}, 
\nonumber 
\end{equation}
and 
\begin{equation}
G({\bf p}, \tau) = {1 \over \beta} \sum_{n'} 
G({\bf p}, i\varepsilon_{n'}) e^{-i\varepsilon_{n'} \tau}, 
\ \ \ \ \ \ \ \ \ \ 
G({\bf p}\leftarrow{\bf p'}, - \tau) = {1 \over \beta} \sum_n 
G({\bf p}\leftarrow{\bf p'}, i\varepsilon_n) e^{i\varepsilon_n \tau}, 
\nonumber 
\end{equation}
we obtain the contributions in Figs.~1-(a) and (b) 
\begin{align}
{\mathbb F}(i\omega_m) = - 2e {1 \over \beta} \sum_n \sum_{\bf p} 
{ {\bf p} \over m } \Big[ 
& G({\bf p+k/2},i\omega_m+i\varepsilon_n) 
G({\bf p+k/2}\leftarrow{\bf p-k/2},i\varepsilon_n) { {\bf p+k/2} \over m } 
\nonumber \\ + 
& G({\bf p+k/2}\leftarrow{\bf p-k/2},i\omega_m+i\varepsilon_n) 
G({\bf p-k/2},i\varepsilon_n) { {\bf p-k/2} \over m } \Big]. 
\nonumber 
\end{align}
}  
in (\ref{def-Phi-e}) are drawn in Fig.~\ref{fig:4-Loop} 
where we only need the first order contributions in ${\bf A}$ 
to obtain the conductivity linear in ${\bf H}$ 
and the second order one has been neglected. 
The product of $j^H_\mu({\bf k})$ and $j^H_\nu(0)$ 
leads to four kinds of terms: 
(i) $j^e_\mu({\bf k}) \cdot j^e_\nu(0)$, 
(ii) $j^e_\mu({\bf k}) \cdot \rho(-{\bf k}) A_\nu$, 
(iii) $\rho(0) A_\mu \cdot j^e_\nu(0)$, and 
(iv) $\rho(0) A_\mu \cdot \rho(-{\bf k}) A_\nu$. 
In the case of (i) two current vertices can be connected by two ways 
as Figs.~\ref{fig:4-Loop}-(a) and (b) 
within the linear order of ${\bf A}$. 
In the cases of (ii) and (iii) 
one of the vertices is already proportional to ${\bf A}$ so that 
the vertices are connected by the electron propagator diagonal in momentum 
as Figs.~\ref{fig:4-Loop}-(c) and (d). 
The case (iv) is not considered here, 
because it is the second order contribution in ${\bf A}$. 

Here $G({\bf p+k/2}\leftarrow{\bf p-k/2}, i\varepsilon_n)$ 
is the Fourier transform of the off-diagonal propagator in momentum variable 
\begin{equation}
G({\bf p+k/2}\leftarrow{\bf p-k/2}, \tau) = - 
\big\langle T_\tau \big\{ c_{{\bf p+k/2}\sigma}(\tau) \ 
                          c_{{\bf p-k/2}\sigma}^\dag \big\} \big\rangle, 
\label{off-diagonal-prop} 
\end{equation}
and evaluated by\footnote{ 
If we write down the rule of Feynman diagram faithfully: 
\begin{equation}
-G({\bf p+k/2}\leftarrow{\bf p-k/2}, i\varepsilon_n) 
\fallingdotseq 
\big[ - G({\bf p}, i\varepsilon_n) \big] \cdot 
\big[ - ( - e {\bf v}_{\bf p} \cdot {\bf A} ) \big] \cdot 
\big[ - G({\bf p}, i\varepsilon_n) \big]. 
\nonumber 
\end{equation}
I prefer the textbook, Lifshitz and Pitaevskii: 
{\it Statistical Physics Part 2} (Pergaman Press, Oxford, 1980), 
because such a faithful description is given concisely. } 
\begin{align}
G({\bf p+k/2}\leftarrow{\bf p-k/2}, i\varepsilon_n) 
& \fallingdotseq 
G({\bf p+k/2}, i\varepsilon_n) \cdot 
\big( - e {\bf v}_{\bf p} \cdot {\bf A} \big) \cdot 
G({\bf p-k/2}, i\varepsilon_n) 
\nonumber \\ 
& \fallingdotseq 
G({\bf p}, i\varepsilon_n) \cdot 
\big( - e {\bf v}_{\bf p} \cdot {\bf A} \big) \cdot 
G({\bf p}, i\varepsilon_n), 
\label{GAG} 
\end{align}
which\footnote{ 
The same relation is obtained 
via the gauge transformation, ${\bf p}\rightarrow{\bf p}-e{\bf A}$. 
For example, in the case of free electron propagator 
\begin{equation}
G_0({\bf p}, i\varepsilon_n) =  
\Big[ i\varepsilon_n - \Big( { {\bf p}^2 \over 2m } - \mu \Big) \Big]^{-1}, 
\nonumber 
\end{equation}
is transformed into 
\begin{equation}
G_0({\bf p}-e{\bf A}, i\varepsilon_n) 
\fallingdotseq G_0({\bf p}, i\varepsilon_n) + 
G_0({\bf p}, i\varepsilon_n) \cdot 
\Big( - {e \over m} {\bf p} \cdot {\bf A} \Big) \cdot 
G_0({\bf p}, i\varepsilon_n). 
\nonumber 
\end{equation}
} 
is the first order perturbation in terms of (\ref{jA-int}) 
in the limit of ${\bf k} \rightarrow 0$. 

In order to obtain the conductivity proportional to the magnetic field 
defined in (\ref{H=rotA}), which is linear in both ${\bf A}$ and ${\bf k}$, 
we have to extract the ${\bf k}$-linear contribution 
from the processes shown in Figs.~\ref{fig:4-Loop}-(a), (b), (c). 
A ${\bf k}$-linear contribution comes from 
the propagators diagonal in momentum variable ${\bf p \pm k/2}$ 
represented by the solid line.\footnote{ 
In the case of free electrons 
the ${\bf k}$-linear contribution of the free propagator 
is easily extracted as 
\begin{equation}
G_0({\bf p \pm k/2}, i\varepsilon_n) 
\fallingdotseq G_0({\bf p}, i\varepsilon_n) + 
G_0({\bf p}, i\varepsilon_n) \cdot 
\Big( \pm { {\bf p} \cdot {\bf k} \over 2m } \Big) \cdot 
G_0({\bf p}, i\varepsilon_n). 
\nonumber 
\end{equation}
} 
It also comes from $j^e_\nu(0)$ 
but does not from $j^e_\mu({\bf k})$.\footnote{ 
In this footnote we use diagonal propagators 
which appear in the  right-hand side of (\ref{GAG}). 
From $j^e_\nu(0)$ vertex we obtain the contribution as 
\begin{equation}
G({\bf p \pm k/2}, i\varepsilon_n+i\omega_\lambda) \cdot 
{ e \over 2m } 
\Big[ \Big( p_\nu \pm { k_\nu \over 2} \Big) 
    + \Big( p_\nu \pm { k_\nu \over 2} \Big) \Big] \cdot 
G({\bf p \pm k/2}, i\varepsilon_n), 
\nonumber 
\end{equation}
which leads to a ${\bf k}$-linear contribution 
\begin{equation}
G({\bf p}, i\varepsilon_n+i\omega_\lambda) \cdot 
\Big( \pm { e \over 2m } k_\nu \Big) \cdot 
G({\bf p}, i\varepsilon_n). 
\nonumber 
\end{equation}
From $j^e_\mu({\bf k})$ vertex we obtain 
\begin{equation}
G({\bf p + k/2}, i\varepsilon_n+i\omega_\lambda) \cdot 
{ e \over 2m } 
\Big[ \Big( p_\mu +{ k_\mu \over 2} \Big) 
    + \Big( p_\mu -{ k_\mu \over 2} \Big) \Big] \cdot 
G({\bf p - k/2}, i\varepsilon_n), 
\nonumber 
\end{equation}
which reduces to a ${\bf k}$-independent contribution 
\begin{equation}
G({\bf p}, i\varepsilon_n+i\omega_\lambda) \cdot 
\Big( { e \over m } p_\mu \Big) \cdot 
G({\bf p}, i\varepsilon_n). 
\nonumber 
\end{equation}
in the limit of ${\bf k} \rightarrow 0$. 

See the footnote for the current-current correlation function 
in (\ref{def-Phi-e}). } 

The ${\bf k}$-linear part $\Phi^e_{(1)}(i\omega_\lambda)$ 
of $\Phi_{xy}^e({\bf k}, i\omega_\lambda)$ 
is the summation of the ${\bf k}$-linear contributions, 
$\Phi_{(a)}$, $\Phi_{(b)}$, $\Phi_{(c)}$, 
which are extracted from the processes 
shown in Fig.~\ref{fig:4-Loop}-(a), (b), (c). 
Since $j^e_x({\bf k})$ leads to the factor $(e/m) p_x$, 
$j^e_y(0)$ to $(e/2m) k_y$, and 
$G({\bf p+k/2}\leftarrow{\bf p-k/2}, i\varepsilon_n)$ to 
$G({\bf p}, i\varepsilon_n) \cdot [ - (e/m) {\bf p} \cdot {\bf A} ] 
\cdot G({\bf p}, i\varepsilon_n)$, 
\begin{equation}
\Phi_{(a)} = -2 T \sum_n \sum_{\bf p} 
{-1 \over 2} \Big( {e \over m} \Big)^3 p_x^2 
G({\bf p}, i\varepsilon_n+i\omega_\lambda) G({\bf p}, i\varepsilon_n)^2 
k_y A_x, 
\label{Phi(a)} 
\end{equation}
where\footnote{ 
The integrand is proportional to $(p_xA_x+p_yA_y+p_zA_z) \cdot p_x$ 
but the terms proportional to $A_y$ and $A_z$ are odd in $p_x$ and 
vanish by the integration over $p_x$. 
In the same manner the integrand in (\ref{Phi(c)}) is proportional to 
$(p_xk_x+p_yk_y+p_zk_z) \cdot p_x$ 
but the terms proportional to $k_y$ and $k_z$ vanish. } 
the product of the fermion-loop factor 
and the spin-degeneracy factor,\footnote{ 
Throughout the series of three Notes the Zeeman splitting is neglected 
so that the spin degrees of freedom only appears as the degeneracy factor 2. } 
-2, has been included. 
Since $j^e_y(0)$ leads to the factor\footnote{ 
This factor is already proportional to $k_y$ 
so that we can put ${\bf k}=0$ for all the propagators 
in the diagrams (a) and (b), 
because we only need the contribution linear in $k_y$. } 
$-(e/2m) k_y$, 
\begin{equation}
\Phi_{(b)} = -2 T \sum_n \sum_{\bf p} 
{1 \over 2} \Big( {e \over m} \Big)^3 p_x^2 
G({\bf p}, i\varepsilon_n+i\omega_\lambda)^2 G({\bf p}, i\varepsilon_n) 
k_y A_x. 
\label{Phi(b)} 
\end{equation}
Since $\rho(-{\bf k})$ leads to the factor $-(e^2/m) A_y$, and 
$G({\bf p+k/2}, i\varepsilon_n+i\omega_\lambda)$ to 
$G({\bf p}, i\varepsilon_n+i\omega_\lambda) 
\cdot [ + (1/2m) {\bf p} \cdot {\bf k} ] 
\cdot G({\bf p}, i\varepsilon_n+i\omega_\lambda)$, and 
$G({\bf p-k/2}, i\varepsilon_n)$ to 
$G({\bf p}, i\varepsilon_n) 
\cdot [ - (1/2m) {\bf p} \cdot {\bf k} ] 
\cdot G({\bf p}, i\varepsilon_n)$,
\begin{align}
\Phi_{(c)} = -2 T \sum_n \sum_{\bf p} \Big[ 
{-1 \over 2} & \Big( {e \over m} \Big)^3 p_x^2 
G({\bf p}, i\varepsilon_n+i\omega_\lambda)^2 G({\bf p}, i\varepsilon_n) 
k_x A_y 
\nonumber \\
+ 
{1 \over 2} & \Big( {e \over m} \Big)^3 p_x^2 
G({\bf p}, i\varepsilon_n+i\omega_\lambda) G({\bf p}, i\varepsilon_n)^2 
k_x A_y \Big]. 
\label{Phi(c)} 
\end{align}
Thus we obtain 
\begin{align}
\Phi_{(a)}+\Phi_{(b)}+\Phi_{(c)} & = 
- {1 \over 2} \Big( {e \over m} \Big)^3 \big( k_x A_y - k_y A_x \big) 
T \sum_n \sum_{\bf p} p_x^2 
\nonumber \\
& \times 
\Big[ 
G({\bf p}, i\varepsilon_n+i\omega_\lambda) G({\bf p}, i\varepsilon_n)^2 
- 
G({\bf p}, i\varepsilon_n+i\omega_\lambda)^2 G({\bf p}, i\varepsilon_n) 
\Big]. 
\label{Phi(a+b+c)} 
\end{align}
If we set ${\bf H}=(0,0,H)$ so that $H=i(k_xA_y-k_yA_x)$, 
(\ref{Phi(a+b+c)}) leads to\footnote{
Eq.~(2.19) in [FEW] obtained for general dispersion 
reduces to (\ref{Phi_1}) for isotropic dispersion 
${\bf v}_{\bf p} = {\bf p}/m$. } 
\begin{align}
\Phi^e_{(1)}(i\omega_\lambda) & = 
- {H \over i} \Big( {e \over m} \Big)^3 
T \sum_n \sum_{\bf p} p_x^2 
\nonumber \\
& \times
\Big[ 
G({\bf p}, i\varepsilon_n+i\omega_\lambda) G({\bf p}, i\varepsilon_n)^2 
- 
G({\bf p}, i\varepsilon_n+i\omega_\lambda)^2 G({\bf p}, i\varepsilon_n) 
\Big]. 
\label{Phi_1} 
\end{align}

\vskip 4mm
\begin{figure}[htbp]
\begin{center}
\includegraphics[width=10.0cm]{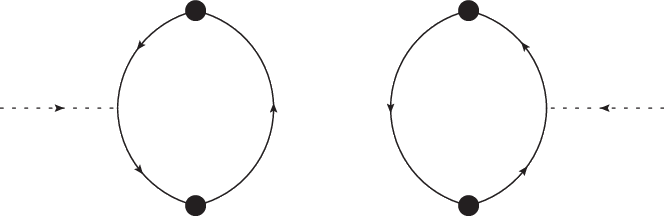}
\caption{Diagrams for a fixed gauge ${\bf A}=(A_x,0,0)$: 
The left diagram corresponds to the one in Fig.~\ref{fig:4-Loop}-(a) 
and the right to Fig.~\ref{fig:4-Loop}-(b). 
The broken line represents the coupling to the magnetic field (\ref{jA-int}). 
(left) The solid line in upward direction is 
$-G({\bf p}, i\varepsilon_n+i\omega_\lambda)$. 
The downward process is the product 
$[-G({\bf p}, i\varepsilon_n)]\cdot[-(-J_x A_x)]
\cdot[-G({\bf p}, i\varepsilon_n)]$. 
The upper black circle is $J_x$ 
and the lower black circle is $(\partial J_y/\partial p_y)k_y/2$. 
(right) The solid line in downward direction is 
$-G({\bf p}, i\varepsilon_n)$. 
The upward process is the product 
$[-G({\bf p}, i\varepsilon_n+i\omega_\lambda)]\cdot[-(-J_x A_x)]
\cdot[-G({\bf p}, i\varepsilon_n+i\omega_\lambda)]$. 
The upper black circle is $J_x$ 
and the lower black circle is $-(\partial J_y/\partial p_y)k_y/2$. 
Here $J_x \equiv (e/m)p_x$ and $\partial J_y/\partial p_y \equiv e/m$. 
}
\label{fig:QP}
\end{center}
\end{figure}
If we choose the gauge ${\bf A}=(A_x,0,0)$, 
we do not have to consider the contribution $\Phi_{(c)}$ 
of the diamagnetic current. 
Such a choice make the calculation simpler.\footnote{ 
The approach by Altshuler, Khmel'nitzkii, Larkin and Lee: 
Phys. Rev. B {\bf 22}, 5142 (1980) 
considers only paramagnetic contributions $\Phi_{(a)}$ and $\Phi_{(b)}$ 
by fixing the gauge. } 
The relevant processes leading to ${\bf H}$-linear contribution 
are summarized in Fig.~\ref{fig:QP}. 
The expressions (\ref{Phi(a)}) and (\ref{Phi(b)}) are rewritten as 
\begin{equation}
- {1 \over 2} \Phi_{(a)} = T \sum_n \sum_{\bf p} 
J_x {\partial J_y \over \partial p_y} {k_y \over 2} 
G(i\varepsilon_n) \big( - J_x A_x \big) G(i\varepsilon_n)
G(i\varepsilon_n+i\omega_\lambda), 
\label{Phi(a)-->} 
\end{equation}
\begin{equation}
- {1 \over 2} \Phi_{(b)} = T \sum_n \sum_{\bf p} 
J_x {\partial J_y \over \partial p_y} \Big( - {k_y \over 2} \Big) 
G(i\varepsilon_n) G(i\varepsilon_n+i\omega_\lambda) 
\big( - J_x A_x \big) G(i\varepsilon_n+i\omega_\lambda), 
\label{Phi(b)-->} 
\end{equation}
where $J_\mu \equiv (e/m) p_\mu$ and 
$G(i\varepsilon_n) \equiv G({\bf p},i\varepsilon_n)$. 
In this case (\ref{Phi_1}) is rewritten as 
\begin{align}
- {1 \over 2} \Phi^e_{(1)}(i\omega_\lambda) & = 
{H \over 2 i} T \sum_n \sum_{\bf p} \big( J_x \big)^2 
{\partial J_y \over \partial p_y} 
\nonumber \\
& \times
\Big[ G(i\varepsilon_n+i\omega_\lambda) G(i\varepsilon_n)^2 
- 
G(i\varepsilon_n+i\omega_\lambda)^2 G(i\varepsilon_n) \Big], 
\label{Phi_1-->} 
\end{align}
where the fermion-loop factor and the spin-degeneracy factor 
are moved to the left-hand side. 
Here $H=-ik_yA_x$. 

In the following I adopt a shortcut\footnote{ 
Using the shortcut formula (\ref{int-XY}) 
the summation (156) in [I] is evaluated as 
\begin{equation}
I^e(\omega+i\delta) - I^e(i\delta) \fallingdotseq 
- \omega \int_{-\infty}^\infty {d\epsilon \over 2\pi i} 
\Big( - {\partial f(\epsilon) \over \partial \epsilon} \Big) 
G^A(\epsilon) G^R(\epsilon), 
\nonumber 
\end{equation}
and (167) as 
\begin{equation}
I^Q(\omega+i\delta) - I^Q(i\delta) \fallingdotseq 
- \omega \int_{-\infty}^\infty {d\epsilon \over 2\pi i} 
\Big( - {\partial f(\epsilon) \over \partial \epsilon} \Big)\ \epsilon \ 
G^A(\epsilon) G^R(\epsilon). 
\nonumber 
\end{equation}
The integral 
\begin{equation}
- \omega \int_{-\infty}^\infty {d\epsilon \over 2\pi i} 
G^A(\epsilon) G^R(\epsilon) = i \omega \tau, 
\nonumber 
\end{equation}
by the residue readily leads to (165) and (171) in [I]. } 
calculation along [FEW] for 
\begin{equation}
I^e(i\omega_\lambda) \equiv - {1 \over \beta} \sum_n 
X(i\varepsilon_n) Y(i\varepsilon_n+i\omega_\lambda), 
\end{equation}
where $X(i\varepsilon_n) \equiv [G({\bf p},i\varepsilon_n)]^a$ 
and $Y(i\varepsilon_n+i\omega_\lambda) 
\equiv [G({\bf p},i\varepsilon_n+i\omega_\lambda)]^b$ 
with $a$ and $b$ being positive integers, 
because the calculation along \S 11 of [I] is cumbersome. 
By repeating the transformation which leads to (159) in [I] 
and neglecting\footnote{ 
Here we employ the propagator (\ref{G^R-G^A}) 
and the location of its pole is explicitly known. 
The integral on $C_1$ with $G^R$ only vanishes 
by choosing a closed contour in which there is no pole. 
The integral on $C_4$ with $G^A$ only also vanishes. 
On the other hand, the integrals on  $C_2$ and $C_3$, 
which contain a pair of $G^R \cdot G^A$ at least, are determined 
by the pole of either $G^R$ or $G^A$. } 
the contributions from $C_1$ and $C_4$ 
we obtain 
\begin{equation}
I^e(i\omega_\lambda) \fallingdotseq 
\int_{-\infty}^\infty {d\epsilon \over 2\pi i} f(\epsilon) 
\Big[ - X^A(\epsilon) Y^R(\epsilon+i\omega_\lambda) 
      + X^A(\epsilon-i\omega_\lambda) Y^R(\epsilon) \Big], 
\end{equation}
where 
the first integrand is the contribution along $C_2$ 
and the second along $C_3$. 
The contours of the integral, $C_1$, $C_2$, $C_3$ and $C_4$, 
are defined in Fig.~5 of [I]. 
To calculate the DC conductivity 
we only need the $\omega$-linear contribution 
\begin{align}
I^e(\omega+i\delta) - I^e(i\delta) & \fallingdotseq 
- \omega \int_{-\infty}^\infty {d\epsilon \over 2\pi i} f(\epsilon) 
\bigg[ X^A(\epsilon) {\partial Y^R(\epsilon) \over \partial \epsilon} 
     + {\partial X^A(\epsilon) \over \partial \epsilon} Y^R(\epsilon) \bigg] 
\nonumber \\ & = 
- \omega \int_{-\infty}^\infty {d\epsilon \over 2\pi i} 
\Big( - {\partial f(\epsilon) \over \partial \epsilon} \Big) 
X^A(\epsilon) Y^R(\epsilon), 
\label{int-XY} 
\end{align}
where we have employed the integration by parts. 
By the same approximation leading to (164) in [I], 
the application of (\ref{int-XY}) to (\ref{Phi_1}) results in 
\begin{align}
& \Phi^e_{(1)}(\omega+i\delta) - \Phi^e_{(1)}(i\delta) 
\nonumber \\
& \sim - i \omega {e^3 \over m} \sum_{\bf p} v_x^2 
\Big( - {\partial f(\xi_{\bf p}) \over \partial \xi_{\bf p}} \Big) 
\int_{-\infty}^\infty {d\epsilon \over 2\pi i} 
\Big[ 
G^A({\bf p}, \epsilon) G^R({\bf p}, \epsilon)^2 
- 
G^A({\bf p}, \epsilon)^2 G^R({\bf p}, \epsilon) 
\Big], 
\end{align}
where 
\begin{equation}
G^R({\bf p}, \epsilon) = {1 \over \epsilon - \xi_{\bf p} + i/2\tau}, 
\ \ \ \ \ \ \ \ \ \ 
G^A({\bf p}, \epsilon) = {1 \over \epsilon - \xi_{\bf p} - i/2\tau}, 
\label{G^R-G^A} 
\end{equation}
which is the analytic continuation 
of the thermal propagator (\ref{thermal-G}). 
The integral over $\epsilon$ is performed 
by evaluating the residue\footnote{ 
\begin{equation}
\int_{-\infty}^\infty {d\epsilon \over 2\pi i} 
\Big[ 
G^A({\bf p}, \epsilon) G^R({\bf p}, \epsilon)^2 
- 
G^A({\bf p}, \epsilon)^2 G^R({\bf p}, \epsilon) 
\Big] = - 2 \tau^2. 
\nonumber 
\end{equation}
} 
so that we finally obtain 
\begin{align}
\sigma_{xy} & \equiv \lim_{\omega \rightarrow 0} \lim_{{\bf k} \rightarrow 0} 
{ 1 \over i\omega } 
\big[ \Phi_{xy}^e({\bf k}, \omega + i\delta) 
- \Phi_{xy}^e({\bf k}, i\delta) \big] 
\nonumber \\ 
& \sim 2 {e^3 \over m} H \sum_{\bf p} 
\Big( - {\partial f(\xi_{\bf p}) \over \partial \xi_{\bf p}} \Big) 
v_x^2 \tau^2, 
\label{Hall-QP} 
\end{align}
which coincides with  the result of the Boltzmann transport (\ref{J^e(H,E)}) 
taking into account that ${\bf H}=(0,0,H)$ and ${\bf E}=(0,E,0)$. 

The expectation value of the heat current ${\bf J}^Q$ 
is expressed as the linear response to electric field 
\begin{equation}
J^Q_\mu({\bf k}, \omega) = 
\sum_\nu {\tilde \alpha}_{\mu\nu}({\bf k}, \omega) 
E_\nu({\bf k} \!=\! 0, \omega), 
\end{equation}
with 
\begin{equation}
{\tilde \alpha}_{\mu\nu}({\bf k}, \omega) = { 1 \over i\omega } 
\big[ \Phi_{\mu\nu}^Q({\bf k}, \omega + i\delta) 
- \Phi_{\mu\nu}^Q({\bf k}, i\delta) \big]. 
\end{equation}
As has been discussed in the above, 
in order to extract the contribution proportional to $H$, 
we only need\footnote{ 
We only need the charge current $j^e_\mu({\bf k})$ 
in the absence of the magnetic field 
to obtain $\Phi_{\mu\nu}^e({\bf k}, \omega + i\delta)$ 
proportional to $H$. } 
the heat current $j^Q_\mu({\bf k})$ 
in the absence of the magnetic field so that 
only the difference between $\Phi^e_{\mu\nu}({\bf k}, i\omega_\lambda)$ 
and $\Phi^Q_{\mu\nu}({\bf k}, i\omega_\lambda)$ 
is the factor\footnote{ 
It should be noted that the factor is proportional to 
the summation of incoming and outgoing frequencies 
$\varepsilon_n + (\varepsilon_n + \omega_\lambda)$. } 
$(i\varepsilon_n + i\omega_\lambda / 2)/e$ 
as in the case of \S 11 in [I]. 

By repeating the above shortcut calculation for 
\begin{equation}
I^Q(i\omega_\lambda) \equiv - {1 \over \beta} \sum_n 
\Big( i\varepsilon_n + {i\omega_\lambda \over 2} \Big) 
X(i\varepsilon_n) Y(i\varepsilon_n+i\omega_\lambda), 
\end{equation}
we obtain 
\begin{equation}
I^Q(\omega+i\delta) \fallingdotseq 
\int_{-\infty}^\infty {d\epsilon \over 2\pi i} f(\epsilon) 
\Big[ - \Big( \epsilon + {\omega \over 2} \Big) 
X^A(\epsilon) Y^R(\epsilon + \omega) 
+ \Big( \epsilon - {\omega \over 2} \Big) 
X^A(\epsilon - \omega) Y^R(\epsilon) \Big]. 
\end{equation}
The $\omega$-linear contribution becomes 
\begin{align}
I^Q(\omega+i\delta) - I^Q(i\delta) & \fallingdotseq 
- \omega \int_{-\infty}^\infty {d\epsilon \over 2\pi i} f(\epsilon) 
\bigg\{ \epsilon \bigg[ 
  X^A(\epsilon) {\partial Y^R(\epsilon) \over \partial \epsilon} 
+ {\partial X^A(\epsilon) \over \partial \epsilon} Y^R(\epsilon) 
\bigg] 
+ X^A(\epsilon) Y^R(\epsilon) \bigg\} 
\nonumber \\ & = 
- \omega \int_{-\infty}^\infty {d\epsilon \over 2\pi i} 
\Big( - {\partial f(\epsilon) \over \partial \epsilon} \Big) 
\ \epsilon \ X^A(\epsilon) Y^R(\epsilon), 
\end{align}
by the integration by parts. 

Thus the ${\bf k}$-linear part $\Phi^Q_{(1)}(i\omega_\lambda)$ 
of $\Phi_{xy}^Q({\bf k}, i\omega_\lambda)$ 
is obtained as 
\begin{align}
& \Phi^Q_{(1)}(\omega+i\delta) - \Phi^Q_{(1)}(i\delta) 
\nonumber \\
& \sim - i \omega {e^2 \over m} \sum_{\bf p} v_x^2 
\Big( - {\partial f(\xi_{\bf p}) \over \partial \xi_{\bf p}} \Big) \xi_{\bf p} 
\int_{-\infty}^\infty {d\epsilon \over 2\pi i} 
\Big[ 
G^A({\bf p}, \epsilon) G^R({\bf p}, \epsilon)^2 
- 
G^A({\bf p}, \epsilon)^2 G^R({\bf p}, \epsilon) 
\Big], 
\end{align}
where we have pulled out the factor 
$ - (\partial f(\epsilon) / \partial \epsilon) \cdot \epsilon$ 
from the integral over $\epsilon$ 
as in the case of \S 11 in [I]. 
Finally we obtain\footnote{ 
The remark in the footnote 40 of [I] 
also applies to (\ref{tilde_alpha_{xy}}). }
\begin{align}
{\tilde\alpha}_{xy} & \equiv 
\lim_{\omega \rightarrow 0} \lim_{{\bf k} \rightarrow 0} 
{ 1 \over i\omega } 
\big[ \Phi_{xy}^Q({\bf k}, \omega + i\delta) 
- \Phi_{xy}^Q({\bf k}, i\delta) \big] 
\nonumber \\ 
& \sim 2 {e^2 \over m} H \sum_{\bf p} 
\Big( - {\partial f(\xi_{\bf p}) \over \partial \xi_{\bf p}} \Big) 
v_x^2 \tau^2 \xi_{\bf p}, 
\label{tilde_alpha_{xy}} 
\end{align}
which coincides with  the result of the Boltzmann transport (\ref{J^Q(H,E)}). 

\section{GL Transport}

The linearized GL transport theory gives\footnote{ 
See the footnote 46 in [I]. 
Since $\tau_2 < 0$ in 3D as discussed in the footnote 61 of [I], 
(\ref{sigma-AL}) for Cooper pairs has the same sign 
as (\ref{Hall-QP}) for electrons. 
It is natural, because $\sigma_{xy}$ is related to the cyclotron motion 
of the charged object 
and both electrons and Cooper pairs carry negative charge. 
By the same reason (\ref{tilde_alpha_{xy}}) for free electrons in 3D 
and (\ref{alpha-AL}) for Cooper pairs have the same sign. 
The same discussion also applies to $\alpha_{xx}$ 
so that (171) in [I] for free electrons in 3D 
and (199) in [I] for Cooper pairs have the same sign. } 
\begin{equation}
\sigma_{xy} 
= {e^2 \over 48 d} {\tau_2 \over \tau_1} {h \over \epsilon^2}, 
\label{sigma-AL} 
\end{equation}
and 
\begin{equation}
\alpha_{xy} = {|e| \over 4 \pi d} {h \over \epsilon}, 
\label{alpha-AL} 
\end{equation}
where $h \equiv 2 |e| H \xi_0^2$. 
The derivation of these results\footnote{ 
The conductivity tensor (\ref{sigma-AL}) and 
the thermo-electric tensor (\ref{alpha-AL}) are given in 
(3.52) and (4.36) of [3] in [I]. 
The contribution of the magnetization current modifies (4.36) into (4.38) 
that is identical to the result in TABLE I of [5] in [I]. } 
shall be discussed 
by the non-perturbational treatment of the magnetic field 
in the third Note. 

\section{Cooper-Pair Transport: AL Process}

\begin{figure}[htbp]
\begin{center}
\includegraphics[width=10.0cm]{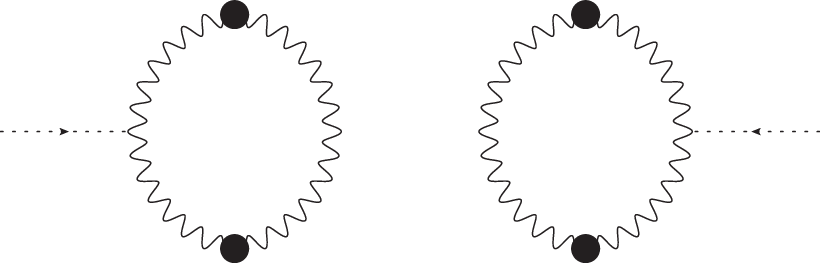}
\caption{AL process for a fixed gauge ${\bf A}=(A_x,0,0)$: 
These diagrams correspond to those in Fig.~\ref{fig:QP}. 
The broken line represents the coupling to the magnetic field (\ref{jA-int}). 
(left) The wavy line in right-side is 
$-L({\bf q}, i\omega_m+i\omega_\lambda)$. 
The left-side process is the product 
$[-L({\bf q}, i\omega_m)]\cdot[-(-{\tilde \Delta}^e_x A_x)]
\cdot[-L({\bf q}, i\omega_m)]$. 
The upper black circle is ${\tilde \Delta}^e_x$ 
and the lower black circle is 
$(\partial {\tilde \Delta}^e_y/\partial q_y)k_y/2$. 
(right) The wavy line in left-side is 
$-L({\bf q}, i\omega_m)$. 
The right-side process is the product 
$[-L({\bf q}, i\omega_m+i\omega_\lambda)]\cdot[-(-{\tilde \Delta}^e_x A_x)]
\cdot[-L({\bf q}, i\omega_m+i\omega_\lambda)]$. 
The upper black circle is ${\tilde \Delta}^e_x$ 
and the lower black circle is 
$-(\partial {\tilde \Delta}^e_y/\partial q_y)k_y/2$. 
Here ${\tilde \Delta}^e_x \equiv 4 e N(0) \xi_0^2 q_x$ 
and $\partial {\tilde \Delta}^e_y/\partial q_y \equiv 4 e N(0) \xi_0^2$. 
}
\label{fig:Cooper}
\end{center}
\end{figure}

The calculations for electrons in \S 3 are 
translated into those for Cooper pairs\footnote{ 
As has been discussed in the footnote 7 of [I], 
$L$ can be identified with the propagator $D_\Delta$ near $T_c$. 
The order-parameter field $\Psi$, (177) in [I], 
is related to the gap function $\Delta$ as 
$\Psi = \sqrt{z} \Delta$ where 
$z = 7 \zeta(3) n / 8 \pi^2 T_c^2$ in 3D. 
\begin{equation}
{\tilde \Delta}_x^e = 4 e N(0) \xi_0^2 q_x 
=2e{3n \over m v_F^2}{7 \zeta(3) v_F^2 \over 48 \pi^2 T_c^2}q_x 
={e^* \over m^*}{7 \zeta(3) n \over 8 \pi^2 T_c^2 }q_x 
=z{e^* \over m^*}q_x,  
\nonumber 
\end{equation}
in 3D. Namely, the current vertex for $\Psi$ is $J_x^e = (e^*/m^*)q_x$ 
in accordance with (185) in [I]. Here $e^*=2e$ and $m^*=2m$. 
For example, 
the value of $z$ is given in (53.23) of [FW] in the footnote 17 of [I]. } 
straightforwardly.\footnote{ 
The perturbational calculation in terms of electron propagators 
shall be given in the Supplement noticed in the footnote 63 of [I]. } 
The contribution of the left diagram in Fig.~\ref{fig:Cooper} is 
\begin{equation}
\Phi_{(a)} = T \sum_m \sum_{\bf q} 
{\tilde \Delta}_x^e 
{\partial {\tilde \Delta}_y^e \over \partial q_y} {k_y \over 2} 
L(i\omega_m) \big( - {\tilde \Delta}_x^e A_x \big) L(i\omega_m) 
L(i\omega_m+i\omega_\lambda), 
\end{equation}
which corresponds to (\ref{Phi(a)-->}) 
and that of the right diagram is 
\begin{equation}
\Phi_{(b)} = T \sum_m \sum_{\bf q} 
{\tilde \Delta}_x^e 
{\partial {\tilde \Delta}_y^e \over \partial q_y} \Big( - {k_y \over 2} \Big) 
L(i\omega_m) L(i\omega_m+i\omega_\lambda) 
\big( - {\tilde \Delta}_x^e A_x \big) L(i\omega_m+i\omega_\lambda), 
\end{equation}
which corresponds to (\ref{Phi(b)-->}) 
where ${\tilde \Delta}_\mu^e \equiv 4 e N(0) \xi_0^2 q_\mu$ and 
$L(i\omega_m) \equiv L({\bf q},i\omega_m)$. 
\vskip 4mm
Thus (\ref{Phi_1-->}) is translated into 
\begin{align}
\Phi^e_{(1)}(i\omega_\lambda) & = 
{H \over 2 i} T \sum_m \sum_{\bf q} \big( {\tilde \Delta}_x^e \big)^2 
{\partial {\tilde \Delta}_y^e \over \partial q_y} 
\nonumber \\
& \times
\Big[ L(i\omega_m+i\omega_\lambda) L(i\omega_m)^2 
- 
L(i\omega_m+i\omega_\lambda)^2 L(i\omega_m) \Big]. 
\end{align}
This result is rewritten as 
\begin{equation}
\Phi^e_{(1)}(i\omega_\lambda) = 32 {H \over i} \big( e \xi_0^2 \big)^3 
\sum_{\bf q} q_x^2 {\tilde I}^e(i\omega_\lambda), 
\end{equation}
with 
\begin{equation}
{\tilde I}^e(i\omega_\lambda) \equiv T \sum_m \Big[ 
{\tilde L}(i\omega_m+i\omega_\lambda)^2 {\tilde L}(i\omega_m) - 
{\tilde L}(i\omega_m+i\omega_\lambda) {\tilde L}(i\omega_m)^2 \Big], 
\end{equation}
where $L(i\omega_m) \equiv - {\tilde L}(i\omega_m) / N(0)$. 
Analytical continuation of this discrete summation becomes 
\begin{equation}
{\tilde I}^e(\omega+i\delta) 
= \int_{-\infty}^\infty {dx \over \pi} n(x) 
\Big\{ \big[ R(+)^2 - A(-)^2 \big] {\rm Im}R 
- \big[ R(+) - A(-) \big] {\rm Im}R^2 \Big\}, 
\end{equation}
where 
\begin{equation}
R = {1 \over \eta - i \tau_0 x}, 
\ \ \ \ \ \ \ \ \ \ 
A = {1 \over \eta + i \tau_0^* x}, 
\end{equation}
\begin{equation}
R(+) = {1 \over \eta - i \tau_0 (x+\omega)}, 
\ \ \ \ \ \ \ \ \ \ 
A(-) = {1 \over \eta + i \tau_0^* (x-\omega)}. 
\end{equation}
Here $\tau_0$ is complex,\footnote{ 
The origin of $\tau_2$ is discussed in the footnote 47 of [I]. 
The perturbational derivation of $\tau_2$ shall be discussed 
in the Supplement noticed in the footnote 63 of [I]. } 
$\tau_0 \equiv \tau_1 + i \tau_2$, 
and $\tau_1 \gg |\tau_2|$. 
The $\omega$-linear contribution is evaluated as 
\begin{align}
{\tilde I}^e(\omega+i\delta) - {\tilde I}^e(i\delta) 
& \fallingdotseq \omega \int_{-\infty}^\infty {dx \over \pi} n(x) 
\Big\{ \Big[ 2R{\partial R \over \partial x} 
          + 2A{\partial A \over \partial x} \Big] {\rm Im}R 
- \Big[ {\partial R \over \partial x} + {\partial A \over \partial x} \Big] 
{\rm Im}R^2 \Big\} 
\nonumber \\ 
& = \omega \int_{-\infty}^\infty {dx \over 2 \pi i} n(x) \big( R-A \big) 
\Big\{ R{\partial R \over \partial x} 
    + A{\partial A \over \partial x} 
    - R{\partial A \over \partial x} 
    - A{\partial R \over \partial x} \Big\}, 
\end{align}
where ${\rm Im}R = (R-A)/2i$ and ${\rm Im}R^2 = (R+A)(R-A)/2i$. 
Employing the high-temperature expansion,\footnote{ 
The statement between (210) and (211) in [I] is insufficient. 
The cut-off frequency $\omega_c$ of the fluctuation propagator, 
(35) in [I], is determined by the condition $\epsilon = \omega_c \tau_0$. 
Since $\tau_0 = \pi / 8T$, $\omega_c/T = 8\epsilon/\pi$ 
so that we can use the high-temperature expansion 
for the integrand with $x < \omega_c$ 
in the limit of $\epsilon \rightarrow 0$. } 
$n(x) \fallingdotseq T/x$, 
\begin{equation}
{\tilde I}^e (\omega+i\delta) - {\tilde I}^e(i\delta) \fallingdotseq 
\omega T \cdot J, 
\end{equation}
with 
\begin{equation}
J \equiv 
\int_{-\infty}^\infty {dx \over 2 \pi i} { R-A \over x } 
\Big\{ R{\partial R \over \partial x} 
    + A{\partial A \over \partial x} 
    - R{\partial A \over \partial x} 
    - A{\partial R \over \partial x} \Big\}. 
\end{equation}
Here we have picked up\footnote{ 
${\rm Im}{\tilde L}^R(x) = ( R - A )/ 2i$. } 
\begin{equation}
{ R-A \over ix } = (\tau_0^* + \tau_0) RA, 
\label{R-A=RA} 
\end{equation}
in the integrand, 
because ${\rm Im}L^R(x)/x$ is proportional to the Lorentz function in $x$ 
and basic quantity. 
Using 
\begin{equation}
{\partial R \over \partial x} = i \tau_0 R^2, 
\ \ \ \ \ \ \ \ \ \ 
{\partial A \over \partial x} = - i \tau_0^* A^2, 
\end{equation}
we obtain\footnote{ 
We have used 
\begin{equation}
\int_{-\infty}^\infty dx 
\Big( R{\partial A \over \partial x} 
    + A{\partial R \over \partial x} \Big) RA = 0. 
\nonumber 
\end{equation}
This is derived by the integration by parts 
noting that 
\begin{equation}
{\partial \over \partial x} \big( RA \big)^2 = 2 RA 
\Big( R{\partial A \over \partial x} 
    + A{\partial R \over \partial x} \Big). 
\nonumber 
\end{equation}
} 
\begin{equation}
J = \int_{-\infty}^\infty {dx \over 2 \pi i} (\tau_0^* + \tau_0) RA 
\big\{ - \tau_0 R^3 + \tau_0^* A^3 \big\}. 
\end{equation}
The integral is evaluated by the residue\footnote{ 
\begin{equation}
\int_{-\infty}^\infty dx R^4 A = 
{2\pi \over \tau_0^*} 
{1 \over \big( 1 + {\tau_0 \over \tau_0^*} \big)^4 } 
{1 \over \eta^4}. 
\nonumber 
\end{equation}
\begin{equation}
\int_{-\infty}^\infty dx R A^4 = 
{2\pi \over \tau_0} 
{1 \over \big( 1 + {\tau_0^* \over \tau_0} \big)^4 } 
{1 \over \eta^4}. 
\nonumber 
\end{equation}
} 
so that 
\begin{equation}
J = - i { \tau_0^* \tau_0 (\tau_0 - \tau_0^*) \over (\tau_0^* + \tau_0)^2 } 
{1 \over \eta^4} 
\fallingdotseq { \tau_2 \over 2 } {1 \over \eta^4}. 
\end{equation}
Therefore 
\begin{equation}
\Phi^e_{(1)}(\omega+i\delta) - \Phi^e_{(1)}(i\delta) \fallingdotseq 
16 {H \over i} e^3 \omega T \tau_2 
\sum_{\bf q} { \xi_0^6 q_x^2 \over (\epsilon + \xi_0^2 {\bf q}^2)^4 }, 
\end{equation}
and this leads to 
\begin{equation}
\sigma_{xy} = 
- 16 H e^3 T \tau_2 
\sum_{\bf q} { \xi_0^6 q_x^2 \over (\epsilon + \xi_0^2 {\bf q}^2)^4 }. 
\end{equation}
In 2D the ${\bf q}$-summation is performed as\footnote{
\begin{equation}
\int_0^\infty dx {x \over (x+\epsilon)^4} 
= \int_0^\infty dx {1 \over (x+\epsilon)^3} 
- \epsilon \int_0^\infty dx {1 \over (x+\epsilon)^4} 
= {1 \over 6} {1 \over \epsilon^2}. 
\nonumber 
\end{equation}
} 
\begin{equation}
\sum_{\bf q} { \xi_0^4 q_x^2 \over (\epsilon + \xi_0^2 {\bf q}^2)^4 } 
= {1 \over 8 \pi d} \int_0^\infty dx {x \over (x + \epsilon)^4} 
= {1 \over 48 \pi d} {1 \over \epsilon^2}. 
\end{equation}
Finally we obtain (\ref{sigma-AL}) 
by using $\tau_1 = \pi / 8T$. 

Only the difference between $\Phi^e_{\mu\nu}({\bf k}, i\omega_\lambda)$ 
and $\Phi^Q_{\mu\nu}({\bf k}, i\omega_\lambda)$ 
is the factor $(i\omega_m + i\omega_\lambda / 2)/2e$ 
as has been discussed in \S 3 so that we readily obtain 
\begin{equation}
\Phi^Q_{(1)}(i\omega_\lambda) = 32 {H \over i} \big( e \xi_0^2 \big)^3 
\sum_{\bf q} q_x^2 {1 \over 2e} {\tilde I}^Q(i\omega_\lambda), 
\end{equation}
with 
\begin{equation}
{\tilde I}^Q(i\omega_\lambda) \equiv T \sum_m 
\Big( i\omega_m + {i\omega_\lambda \over 2} \Big) \Big[ 
{\tilde L}(i\omega_m+i\omega_\lambda)^2 {\tilde L}(i\omega_m) - 
{\tilde L}(i\omega_m+i\omega_\lambda) {\tilde L}(i\omega_m)^2 \Big]. 
\end{equation}
Analytical continuation of this discrete summation becomes\footnote{
Previously published formulae, 
(35) in [Uss], the formula between (10.35) and (10.36) in [2] of [I] 
and (7) in [LNV], 
differ significantly from ours (\ref{I1+I2}). 

[Uss] $\equiv$ Ussishkin: Phys. Rev. B {\bf 68}, 024517 (2003). 

[LNV] $\equiv$ Levchenko, Norman and Varlamov: 
Phys. Rev. B {\bf 83}, 020506 (2011). } 
\begin{align}
{\tilde I}^Q(\omega+i\delta) 
& = \int_{-\infty}^\infty {dx \over \pi} n(x) \Big( x + {\omega \over 2} \Big) 
\Big\{ R(+)^2 \cdot {\rm Im}R - R(+) \cdot {\rm Im}R^2 \Big\} 
\nonumber \\ 
& + \int_{-\infty}^\infty {dx \over \pi} n(x) \Big( x - {\omega \over 2} \Big) 
\Big\{ {\rm Im}R^2 \cdot A(-) - {\rm Im}R \cdot A(-)^2 \Big\} 
\nonumber \\ 
& \equiv {\tilde I}_1^Q(\omega+i\delta) + {\tilde I}_2^Q(\omega+i\delta), 
\label{I1+I2} 
\end{align}
where 
\begin{equation}
{\tilde I}_1^Q(\omega+i\delta) = 
\int_{-\infty}^\infty {dx \over \pi} n(x) \ x \ 
\Big\{ \big[ R(+)^2 - A(-)^2 \big] {\rm Im}R 
- \big[ R(+) - A(-) \big] {\rm Im}R^2 \Big\}, 
\end{equation}
\begin{equation}
{\tilde I}_2^Q(\omega+i\delta) = {\omega \over 2} 
\int_{-\infty}^\infty {dx \over \pi} n(x) 
\Big\{ \big[ R(+)^2 + A(-)^2 \big] {\rm Im}R 
- \big[ R(+) + A(-) \big] {\rm Im}R^2 \Big\}. 
\end{equation}
In the following the imaginary part of $\tau_0$ is neglected: 
$\tau_0^* = \tau_0$ or $\tau_2 = 0$. 
Employing the high-temperature expansion, $n(x) \fallingdotseq T/x$, 
we obtain\footnote{ 
If we put $R \equiv u(x) + i v(x)$ and $A \equiv u(x) - i v(x)$, 
$u(x)$ is even: $u(-x) = u(x)$ and $v(x)$ is odd: $v(-x) = - v(x)$ in $x$. 
Namely $R+A$ is even and $R-A$ is odd. 
Therefore the integrand $(R+A)(R-A)^3$ is odd 
so that the integral in (\ref{null}) vanishes. 

\begin{equation}
\int_{-\infty}^\infty dx R^2 A^2 = 
{\pi \over 2} {1 \over \tau_0} {1 \over \eta^3}. 
\nonumber 
\end{equation}
} 
\begin{align}
{\tilde I}_1^Q(\omega+i\delta) - {\tilde I}_1^Q(i\delta) 
& \fallingdotseq {T \over \pi} \omega \int_{-\infty}^\infty dx 
\Big\{ \Big[ 2R{\partial R \over \partial x} 
          + 2A{\partial A \over \partial x} \Big] {\rm Im}R 
- \Big[ {\partial R \over \partial x} + {\partial A \over \partial x} \Big] 
{\rm Im}R^2 \Big\} 
\nonumber \\ 
& = {T \over \pi} \omega \int_{-\infty}^\infty dx 
{\tau_0 \over 2} \big( R+A \big) \big( R-A \big)^3 = 0, 
\label{null} 
\end{align}
\begin{align}
{\tilde I}_2^Q(\omega+i\delta) - {\tilde I}_2^Q(i\delta) 
& \fallingdotseq {T \over 2\pi} \omega \int_{-\infty}^\infty {dx \over x} 
\Big\{ \Big( R^2 + A^2 \Big) {\rm Im}R 
- \Big( R + A \Big) {\rm Im}R^2 \Big\} 
\nonumber \\ 
& = {T \over 2\pi} \omega \int_{-\infty}^\infty dx 
\big( -2 \tau_0 \big) R^2 A^2 
\nonumber \\ 
& = - {T \over 2} \omega {1 \over \eta^3}. 
\end{align}
Therefore 
\begin{equation}
\Phi^Q_{(1)}(\omega+i\delta) - \Phi^Q_{(1)}(i\delta) \fallingdotseq 
8 H e^2 i \omega T 
\sum_{\bf q} { \xi_0^6 q_x^2 \over (\epsilon + \xi_0^2 {\bf q}^2)^3 }. 
\end{equation}
In 2D the ${\bf q}$-summation is performed as\footnote{
\begin{equation}
\int_0^\infty dx {x \over (x+\epsilon)^3} 
= \int_0^\infty dx {1 \over (x+\epsilon)^2} 
- \epsilon \int_0^\infty dx {1 \over (x+\epsilon)^3} 
= {1 \over 2} {1 \over \epsilon}. 
\nonumber 
\end{equation}
} 
\begin{equation}
\sum_{\bf q} { \xi_0^4 q_x^2 \over (\epsilon + \xi_0^2 {\bf q}^2)^3 } 
= {1 \over 8 \pi d} \int_0^\infty dx {x \over (x + \epsilon)^3} 
= {1 \over 16 \pi d} {1 \over \epsilon}. 
\end{equation}
Finally we obtain (\ref{alpha-AL}) 
via 
\begin{equation}
{\tilde \alpha}_{xy} = T 
{e^2 \over 2 \pi d} H \xi_0^2 {1 \over \epsilon}. 
\end{equation}

The calculations in the absence of the magnetic field 
are also performed in the same manner. 
The integral (207) in [I] with real $\tau_0$ is evaluated as\footnote{ 
\begin{equation}
\int_{-\infty}^\infty dx R^3 A = \int_{-\infty}^\infty dx R A^3 = 
{\pi \over 4} {1 \over \tau_0} {1 \over \eta^3}. 
\nonumber 
\end{equation}
} 
\begin{align}
I^e(\omega+i\delta) - I^e(i\delta) 
& \fallingdotseq {1 \over N(0)^2} {T \over \pi} \omega 
\int_{-\infty}^\infty dx 
\Big[ {\partial R \over \partial x} - {\partial A \over \partial x} \Big] 
{ R-A \over 2 i x } 
\nonumber \\ 
& = {1 \over N(0)^2} {T \over \pi} i \omega \tau_0^2 
\int_{-\infty}^\infty dx \big( R^2 + A^2 \big) RA 
\nonumber \\ 
& = {T \over 2 N(0)^2} i \omega \tau_0 {1 \over \eta^3}. 
\end{align}
The integral (216) in [I] with complex $\tau_0$ is evaluated as\footnote{ 
Noting that 
\begin{equation}
{\partial \over \partial x} \big( R - A \big)^2 
= 2 \big( R - A \big) 
\Big( {\partial R \over \partial x} 
    - {\partial A \over \partial x} \Big), 
\nonumber 
\end{equation}
the integral in the first line of (\ref{(218)}) is shown to vanish 
by integration by parts. 
The integral in the second line is evaluated using (\ref{R-A=RA}) and 
\begin{equation}
\int_{-\infty}^\infty dx R^2 A = 
{2\pi \over \tau_0^*} 
{1 \over \big( 1 + {\tau_0 \over \tau_0^*} \big)^2 } 
{1 \over \eta^2}, 
\nonumber 
\end{equation}
\begin{equation}
\int_{-\infty}^\infty dx R A^2 = 
{2\pi \over \tau_0} 
{1 \over \big( 1 + {\tau_0^* \over \tau_0} \big)^2 } 
{1 \over \eta^2}. 
\nonumber 
\end{equation}
} 
\begin{align}
I^Q(\omega+i\delta) - I^Q(i\delta) 
& \fallingdotseq {1 \over N(0)^2} {T \over \pi} \omega 
\int_{-\infty}^\infty dx 
\Big[ {\partial R \over \partial x} - {\partial A \over \partial x} \Big] 
{ R-A \over 2 i } 
\nonumber \\ 
& + {1 \over N(0)^2} {T \over \pi} {\omega \over 2} 
\int_{-\infty}^\infty dx \big( R - A \big) { R-A \over 2 i x }, 
\label{(218)} 
\end{align}
to give (220) in [I]. 

\section{Remarks}

The sections of {\bf Exercise} and {\bf Acknowledgements} 
are common to [I] so that I do not repeat here. 
Some typographic errors in [I] are listed below. 

\begin{list}{}{}
\item[$\bullet$] The thermodynamic relation in p.~11 should be 
$dQ = dU - \mu dN$. 
\item[$\bullet$] The electric field in (68) and (78) should be 
$E_\nu({\bf k} \!=\! 0, \omega)$. 
\item[$\bullet$] The factor $e$ in the first line of (73) should be 
$e/m$. 
\item[$\bullet$] In Fig.~5 
the subscript $\lambda$ for the lower cut $C_\lambda$ 
has dropped by the font-error at {\boldmath$\mathsf{arXiv}$}. 
\item[$\bullet$] In the footnote 47 
the first term in the right-hand side of the complex GL equation should be 
$(1+ic_0){\tilde \Psi}$. 
\end{list}

(These errors have been fixed in v2.)

\vskip 40pt
\noindent{\bf\Large Appendix}
\vskip 20pt

The linear response theory for the DC Hall conductivity 
of the Dirac fermion in 2+1 space-time dimensions is reviewed. 
One focus is the Chern-Simons effective action for the gauge field. 
Another is the exact formula by Ishikawa and Matsuyama. 

\vskip 40pt
\noindent{\bf\Large A1. Introduction}
\vskip 20pt

The topological nature of the DC Hall conductivity in 2+1 space-time dimensions 
was intensively discussed in 1980's in the context of the quantum Hall effect. 
The discussion based on the Dirac fermion was a major topic in those days. 
In the study of topological insulators 
there is a revival of interest 
in the description by the Dirac fermion in these days. 
Thus this review of the linear response theory 
for the DC Hall conductivity of the Dirac fermion 
might be useful to beginners in the 21st century. 

In the section 2 
the details of the perturbational derivation of the Chern-Simons effective action 
for the gauge field at zero temperature are given. 
 
In the section 3 
a brief discussion on the exact formula at zero temperature 
by Ishikawa and Matsuyama is given. 

\vskip 40pt
\noindent{\bf\Large A2. Chern-Simons action}
\vskip 20pt

We consider the coupled system of the Dirac fermion and the gauge field 
in 2+1 space-time dimensions. 
By tracing out the fermion field\footnote{
The outline is given, for example, 
in~\cite{D} or~\cite{F}. 
We also derived  the Chern-Simons effective action [NKF] 
in the context of the chiral spin liquid and the anyon superconductivity. 
[NKF] $\equiv$ 
Narikiyo, Kuboki, Fukuyama: J. Phys. Soc. Jpn. {\bf 59}, 2443 (1990).} 
we obtain the Chern-Simons effective action for the gauge field.
 
The Lagrangian density for the Dirac fermion is given by 
\begin{equation}
{\cal L} = {\bar \psi} (i \gamma_\mu \partial^\mu - m) \psi, 
\end{equation}
where
${\bar \psi} = \psi^\dagger \gamma^0$ 
and\footnote{
For example, see the section 1 of [D]. 
Since 
\begin{equation}
A_0=A^0, \ \ \ A_1=-A^1, \ \ \ A_2=-A^2, \ \ \ A_3=-A^3, 
\nonumber 
\end{equation}
then 
\begin{equation}
A_\mu B^\mu = A^\mu B_\mu = A_0B_0 - A_1B_1 - A_2B_2 -A_3B_3, 
\nonumber 
\end{equation}
in 3+1 space-time dimensions 
where $x^\mu=(t,{\bf x})$ and $\partial_\mu = \partial/\partial x^\mu$. 
[D] $\equiv$ Dirac: {\it General Theory of Relativity} (Wiley, New York, 1975).} 
$\gamma_\mu \partial^\mu = 
 \gamma_0 \partial_0 - \gamma_1 \partial_1 - \gamma_2 \partial_2$.   

Since 
\begin{equation}
\gamma_0 = \sigma_z, \ \ \ \gamma_1 = i \sigma_x, \ \ \ \gamma_2 = i \sigma_y, 
\end{equation}
the properties of the gamma matrices\footnote{
The basic properties of the Pauli matrices are: 
${\rm tr}[\sigma_x]={\rm tr}[\sigma_y]={\rm tr}[\sigma_z]=0$, 
$\sigma_x^2=\sigma_y^2=\sigma_z^2=I$, 
$\sigma_x\sigma_y=-\sigma_y\sigma_z=i\sigma_z$ and 
$\sigma_x\sigma_y\sigma_z=iI$. 
Thus 
$\gamma_0\gamma_0=-\gamma_1\gamma_1=-\gamma_2\gamma_2=I$ and 
$\gamma_\mu\gamma_\nu=-\gamma_\nu\gamma_\mu$ for $\mu\neq\nu$. 
Since $\gamma_0\gamma_1\gamma_2 = -iI$, 
we obtain ${\rm tr}[\gamma_0\gamma_1\gamma_2] = -2i$. 
Since $\gamma_\lambda\gamma_\mu\gamma_\mu = \pm \gamma_\lambda$, 
we obtain ${\rm tr}[\gamma_\lambda\gamma_\mu\gamma_\mu] = 0$. 
With these results and the anti-symmetric property of gamma matrices 
the trace is summarized as 
${\rm tr}[\gamma_\lambda\gamma_\mu\gamma_\nu] = 
-2i\epsilon_{\lambda\mu\nu}$.}
are those of the Pauli matrices. 
 
The free propagator\footnote{
Assuming the translational invariance 
the propagator is defined as 
\begin{equation}
iG(x-y) 
= \big\langle 0 \big| T \big\{ \psi(x){\bar \psi}(y) \big\} \big| 0 \big\rangle. 
\nonumber 
\end{equation}
The Fourier transform is defined as 
\begin{equation}
G(x-y) = \int{d^3p \over (2\pi)^3} G(p) \exp [-ip(x-y)]. 
\end{equation}
}
of the Dirac fermion is given by the matrix 
\begin{equation}
G(p) = {1 \over {\slashed p} - m + i \delta} 
     = {{\slashed p} + m \over p^2 - m^2 + i \delta}, 
\end{equation}
where\footnote{
Applying 
$\gamma_0\gamma_0=I$, $\gamma_1\gamma_1=-I$, $\gamma_2\gamma_2=-I$ and 
$\gamma_\mu\gamma_\nu=-\gamma_\nu\gamma_\mu$ for $\mu\neq\nu$ to 
$p^2=(\slashed p)^2=
(p_0\gamma_0 - p_1\gamma_1 - p_2\gamma_2)
(p_0\gamma_0 - p_1\gamma_1 - p_2\gamma_2)$ we obtain 
$p^2 = (p_0^2 - p_1^2 - p_2^2) I$.} 
$\slashed p = \gamma_\mu p^\mu 
            = p_0\gamma_0 - p_1\gamma_1 - p_2\gamma_2$, 
$p^2 = p_0^2 - p_1^2 - p_2^2$ and $\delta = +0$. 
The component proportional to $\gamma_0$ is\footnote{
The component is equivalent to 
that in the Nambu representation for superconductivity, 
(7-52) in [S]. 
Here 
$ E_{\bf p} - i \delta' \fallingdotseq \sqrt{\rho} e^{-i\theta/2}$ 
with 
$ E_{\bf p}^2 - i \delta = \rho e^{-i\theta} $. 
[S] $\equiv$ Schrieffer: {\it Theory of Superconductivity} 
(Benjamin, Reading Massachusetts, 1964).} 
\begin{equation}
{1 \over 2} \Big( {1 \over p_0 - E_{\bf p} + i \delta'} 
                + {1 \over p_0 + E_{\bf p} - i \delta'}  \Big), 
\end{equation}
where 
$ E_{\bf p} = \sqrt{p_1^2 + p_2^2 + m^2} $ with ${\bf p}=(p_1,p_2)$. 
In my notation $m>0$ and $E_{\bf p}>0$. 

The coupling between the fermion and the gauge fields 
is given by the interaction Lagrangian \footnote{
See, for example, the section 15.2 of [BD]. 
[BD] $\equiv$ Bjorken and Drell: {\it Relativistic Quantum Fields} 
(McGraw-Hill, New York, 1965).} 
\begin{equation}
{\cal L}_{\rm int} = - e j_\mu(x) A^\mu(x), 
\end{equation}
where\footnote{
Since we are interested in the electron transport, $e<0$. 
The electric current $J_\mu$ is related to the particle current $j_\mu$ 
by $J_\mu = - e j_\mu$. 
The particle current satisfies the conservation law $\partial^\mu j_\mu = 0$. 
See, for example, the section 3.4 of [PS]. 
[PS] $\equiv$ Peskin and Schroeder: {\it An Introduction to Quantum Field Theory} 
(Westview, Boulder, 1995).} 
\begin{equation}
j_\mu(x) = {\bar \psi}(x) \gamma_\mu \psi(x). 
\end{equation}

The effective action $S$ for the gauge field is obtained as 
\begin{equation}
i S = {i^2 \over 2!} \int{d^3 x} \int{d^3 y} A^\mu(x) 
\big\langle 0 \big| T \big\{ J_\mu(x)J_\nu(y) \big\} \big| 0 \big\rangle A^\nu(y), 
\label{2nd-order-S} 
\end{equation}
within the second order perturbation\footnote{
See, for example, the section 5-1-5 of [IZ] 
for the discussion on the generating functional $Z[A]$. 
Suppressing the super/subscript 
\begin{equation}
Z[A] = \big\langle 0 \big| 
T \exp\big[ i \int d^3 x J(x) A(x) \big] \big| 0 \big\rangle 
= \sum_{n=0}^\infty { i^n \over n! } 
\int d^3 x_1 \cdot\cdot\cdot d^3 x_n G(x_1, \cdot\cdot\cdot, x_n)
A(x_1) \cdot\cdot\cdot A(x_n), 
\nonumber 
\end{equation}
where the Green function is defined by 
\begin{equation}
G(x_1, \cdot\cdot\cdot, x_n) = 
\big\langle 0 \big| 
T \big\{ J(x_1) \cdot\cdot\cdot J(x_n) \big\} \big| 0 \big\rangle. 
\nonumber 
\end{equation}
Introducing the effective action $S$ as $Z[A]=\exp(iS)$ 
\begin{equation}
i S = \sum_{n=1}^\infty { i^n \over n! } 
\int d^3 x_1 \cdot\cdot\cdot d^3 x_n G_{\rm c}(x_1, \cdot\cdot\cdot, x_n)
A(x_1) \cdot\cdot\cdot A(x_n), 
\nonumber 
\end{equation}
where $G_{\rm c}$ is the connected Green function. 
The expectation value of the current in the vacuum 
$ \langle 0 | J_\mu(x) | 0 \rangle $ vanishes 
so that we obtain (\ref{2nd-order-S}) within the second order perturbation. 
[IZ] $\equiv$ Itzykson and Zuber: {\it Quantum Field Theory} 
(McGraw-Hill, New York, 1980).}
in $ {\cal L}_{\rm int}$. 
Assuming the translational invariance, 
\begin{equation}
\big\langle 0 \big| T \big\{ J_\mu(x)J_\nu(y) \big\} \big| 0 \big\rangle 
= i\Pi_{\mu\nu}(x-y),  
\end{equation}
and introducing the Fourier transform 
\begin{equation}
\Pi_{\mu\nu}(x-y) = \int{d^3q \over (2\pi)^3} \Pi_{\mu\nu}(q) \exp [-iq(x-y)],  
\end{equation}
$S$ is expressed as\footnote{
The Fourier transform 
\begin{equation}
A^\mu(x) = \int{d^3p \over (2\pi)^3} A^\mu(p) \exp (-ipx), \ \ \ 
A^\nu(y) = \int{d^3p' \over (2\pi)^3} A^\nu(p') \exp (-ip'y), 
\nonumber  
\end{equation}
and 
the integral representation of the delta function 
\begin{equation}
\delta(p+q) = \int{d^3x \over (2\pi)^3} \exp [-i(p+q)x], \ \ \ 
\delta(p'-q) = \int{d^3y \over (2\pi)^3} \exp [-i(p'-q)y],
\nonumber  
\end{equation}
are employed. Here $px = p_\mu x^\mu$.} 
\begin{equation}
S = - {1 \over 2} \int {d^3 q \over (2\pi)^3}
    A^\mu(-q) \Pi_{\mu\nu}(q) A^\nu(q), 
\label{action} 
\end{equation}
where
\begin{equation}
\Pi_{\mu\nu}(q) = i e^2 \int {d^3 p \over (2\pi)^3} {\rm tr} 
\big[\gamma_\mu G\big(p+{q \over 2}\big) \gamma_\nu G\big(p-{q \over 2}\big)\big], 
\label{Pi} 
\end{equation}
in the perturbational calculation.\footnote{
In the faithful representation 
\begin{equation}
i \Pi_{\mu\nu}(q) = (-ie)^2 (-1) \int {d^3 p \over (2\pi)^3} {\rm tr} 
\big[\gamma_\mu iG\big(p+{q \over 2}\big) \gamma_\nu iG\big(p-{q \over 2}\big)\big], 
\nonumber 
\end{equation}
which is equivalent to (19.10) of [BD]. 
I recommend such a faithful representation 
in the footnote 13 of {\ttfamily arXiv:1203.0127}. 
The virtue of it 
will be also shown in the forthcoming Supplement 
for the perturbational calculation of the current vertex for Cooper pairs.}  

The Chern-Simons action is the $q$-linear contribution of (\ref{action}). 
The integrand of (\ref{Pi}) is 
\begin{equation}
{
{\rm tr} 
\big[\gamma_\mu ({\slashed p}+{\slashed q}/2+m) 
     \gamma_\nu ({\slashed p}-{\slashed q}/2+m)\big] 
\over
\big[(p+q/2)^2-m^2+i\delta \big]
\big[(p-q/2)^2-m^2+i\delta \big]
}, 
\end{equation}
The $q$-linear contributions from $(p+q/2)^2$ and $(p-q/2)^2$ cancel out.\footnote{
Since we adopt the symmetric representation $G(p+q/2)G(p-q/2)$, 
the extraction of the $q$-linear contribution is easy. 
If we adopt $G(p+q)G(p)$, it is not so easy. 
This is a virtue of the symmetric representation. 
Another virtue is discussed in the footnote 6 of {\ttfamily arXiv:1112.1513}.}
The $q$-linear contribution from the trace over the gamma matrix is\footnote{
If $\nu \neq \lambda$, 
then $\gamma_\nu \gamma_\lambda = - \gamma_\lambda \gamma_\nu$. 
If $\nu = \lambda$, 
then $\gamma_\nu \gamma_\lambda = \gamma_\lambda \gamma_\nu = \pm I$ 
so that (\ref{trace3}) is satisfied by 
${\rm tr}\big[\gamma_\mu \gamma_\lambda \gamma_\nu \big] = 0$.} 
\begin{equation}
{m \over 2}{\rm tr}\big[\gamma_\mu {\slashed q} \gamma_\nu \big] - 
{m \over 2}{\rm tr}\big[\gamma_\mu \gamma_\nu {\slashed q} \big] = 
m {\rm tr}\big[\gamma_\mu \gamma_\lambda \gamma_\nu \big] q^\lambda. 
\label{trace3} 
\end{equation}
Thus\footnote{
${\rm tr}\big[\gamma_\mu \gamma_\lambda \gamma_\nu \big]  
= - 2 i \epsilon_{\mu\lambda\nu}$.} 
\begin{equation}
S = - {i \over 2} \kappa \int {d^3 q \over (2\pi)^3}
    A^\mu(-q) q^\lambda A^\nu(q) \epsilon_{\mu\lambda\nu}, 
\label{S-q} 
\end{equation}
where 
\begin{equation}
\kappa = e^2 \int {d^3 p \over (2\pi)^3} 
{-2im \over ( p^2-m^2+i\delta )^2} 
= -2ime^2 \int_0^\infty {|{\bf p}|d|{\bf p}| \over 2\pi} I(|{\bf p}|), 
\end{equation}
with  
\begin{equation}
I(|{\bf p}|) = \int_{-\infty}^\infty {d p_0 \over 2\pi} 
{1 \over 
( p_0 - E_{\bf p} + i\delta' )^2
( p_0 + E_{\bf p} - i\delta' )^2}, 
\end{equation}
and 
$ |{\bf p}| = \sqrt{p_1^2+p_2^2} $. 
The integral $I(|{\bf p}|)$ is evaluated by the contour 
consisting of the real axis 
and the semi-circle with infinitely large radius 
in the upper half\footnote{
Such a choice properly picks up 
the contribution of the occupied state as seen in the section 9 of [LP]. 
See also Fig. 3.2 of [NO]. 
[LP] $\equiv$ Lifshitz and Pitaevskii: {\it Statistical Physics} Part2 
(Pergamon, Oxford, 1980). 
[NO] $\equiv$ Negele and Orland: {\it Quantum Many-Particle Systems} 
(Perseus, Cambridge Massachusetts, 1988).} 
of the complex $p_0$-plane 
and results in 
\begin{equation}
I(|{\bf p}|) = { i \over 4 } E_{\bf p}^{-3}. 
\end{equation}
As the contribution from the occupied branch\footnote{
The chemical potential $\mu$ is zero in this case 
so that the state with $\epsilon_{\bf p}<0$ is occupied 
where
$\epsilon_{\bf p} = E_{\bf p}$ or $\epsilon_{\bf p} = - E_{\bf p}$.} 
$-E_{\bf p}$ 
we obtain 
\begin{equation}
\kappa = {me^2 \over 4\pi} 
\int_{-\infty}^{-m} {-\epsilon_{\bf p} d\epsilon_{\bf p} \over (-\epsilon_{\bf p})^3} 
= {e^2 \over 4\pi}. 
\label{sigma_H} 
\end{equation}

With this $\kappa$ 
the effective action $S$ in the coordinate representation\footnote{
The Fourier transform 
\begin{equation}
A^\mu(-q) = \int{d^3y} A^\mu(y) \exp (-iqy), \ \ \ 
A^\nu(q) = \int{d^3x} A^\nu(x) \exp (iqx), 
\nonumber  
\end{equation}
and 
the integral representation of the delta function 
\begin{equation}
\delta(y-x) = \int{d^3q \over (2\pi)^3} \exp [-iq(y-x)], 
\nonumber  
\end{equation}
are employed. Here $qx = q_\mu x^\mu$ and 
\begin{equation}
q^\lambda \exp (iqx) = - i \partial^\lambda \exp (iqx). 
\nonumber  
\end{equation}
}
becomes 
\begin{equation}
S = {\kappa \over 2} \int {d^3 x}
    A^\mu(x) \partial^\lambda A^\nu(x) \epsilon_{\mu\lambda\nu}, 
\label{S-x} 
\end{equation}
where 
$\partial^\lambda = \partial/\partial x_\lambda$. 

The expectation value\footnote{
Employing the generating functional $Z[A]$ in the footnote for (\ref{2nd-order-S}) 
the expectation value $\overline{J(x)}$ is given as 
\begin{equation}
\overline{J(x)} = { 1 \over Z[A] } 
\big\langle 0 \big| 
T \big\{ J(x) \exp\big[ i \int d^3 y J(y) A(y) \big] \big\}\big| 0 \big\rangle, 
\nonumber 
\end{equation}
Since 
\begin{equation}
i J(x) \exp\big[ i \int d^3 y J(y) A(y) \big] 
= {\delta \over \delta J(x)} \exp\big[ i \int d^3 y J(y) A(y) \big], 
\nonumber 
\end{equation}
and 
$ i S = \ln Z[A] $, 
we obtain 
\begin{equation}
\overline{J(x)} = {\delta S \over \delta A(x)}. 
\nonumber 
\end{equation}
} 
of the electric current is given as\footnote{
See, for example, the section 1-1-2 of [IZ] 
for the description of the electromagnetic field. 
Using the same convention 
$-J_1=J^1=J_x$ and $E^2=E_y$. 
The parts of the integrand in (\ref{S-x}) which contain $A^1$ are 
\begin{equation}
A^1 \partial^2 A^0 \epsilon_{120}, \ \ \ 
A^1 \partial^0 A^2 \epsilon_{102}, \ \ \ 
A^0 \partial^2 A^1 \epsilon_{021}, \ \ \ 
A^2 \partial^0 A^1 \epsilon_{201}.
\nonumber  
\end{equation}
The summation of the first and second becomes 
\begin{equation}
A^1 \big( \partial^2 A^0 - \partial^0 A^2 \big).
\nonumber  
\end{equation}
The summation of the third and fourth 
becomes the same after the integration by parts.} 
\begin{equation}
\overline{J_1(x)} = {\delta S \over \delta A^1(x)} 
= \kappa \big( \partial^2 A^0 - \partial^0 A^2 \big) = \kappa E^2. 
\end{equation}
Namely, 
\begin{equation}
\overline{J^1(x)} = \sigma_{xy} E^2, 
\end{equation}
with the DC Hall conductivity\footnote{
The sign of $\sigma_{xy}$ is consistent 
with the negative Hall coefficient for the free electron. 
The absolute value is half of the free fermion value. 
The interpretation of the half value 
is given, for example, in the section 16.3.3 of \cite{F}.} 
\begin{equation}
\sigma_{xy} = - {e^2 \over 4\pi}. 
\end{equation}

In the presence of the chemical potential $\mu$ 
the Lagrangian density becomes\footnote{
This shift is the same as that in (6.1) of \cite{IM}.} 
\begin{equation}
{\cal L} + \mu \psi^\dagger \psi = {\cal L} + \mu {\bar \psi} \gamma^0 \psi, 
\end{equation}
since 
$(\gamma_0)^2 = I$. 
Then the pole of the propagator in the complex $p_0$-plane 
is located at\footnote{
The analyticity is specified as (9.9) in [LP].} 
\begin{equation}
p_0 + \mu = \epsilon_{\bf p} - i \delta' {\rm sign}(\epsilon_{\bf p} - \mu), 
\end{equation}
where
$\epsilon_{\bf p} = E_{\bf p}$ or $\epsilon_{\bf p} = - E_{\bf p}$. 
We only pick up the contribution of the pole in the upper plane. 

When $ -m < \mu < m$, 
the branch $-E_{\bf p}$ is fully occupied so that 
$\kappa$ ie equal to (\ref{sigma_H}). 

When $ \mu < -m $, 
the branch $-E_{\bf p}$ is partially occupied so that 
\begin{equation}
\kappa = {me^2 \over 4\pi}
\int_{-\infty}^{\mu} {-\epsilon_{\bf p}d\epsilon_{\bf p} \over (-\epsilon_{\bf p})^3} 
= {e^2 \over 4\pi}{m \over |\mu|}. 
\end{equation}

When $ m < \mu $, 
the branch $-E_{\bf p}$ is fully occupied and 
the branch $+E_{\bf p}$ is partially occupied\footnote{
The pole of the branch $-E_{\bf p}$ 
has the contribution $iE_{\bf p}^{-3}/4$ to $I(|{\bf p}|)$ as mentioned above. 
On the other hand, 
the pole of the branch $+E_{\bf p}$ has the contribution $-iE_{\bf p}^{-3}/4$.}
so that 
\begin{equation}
\kappa = {me^2 \over 4\pi} \biggl( 
\int_{-\infty}^{-m} {-\epsilon_{\bf p} d\epsilon_{\bf p} \over (-\epsilon_{\bf p})^3} - 
\int_{m}^{\mu} {\epsilon_{\bf p} d\epsilon_{\bf p} \over \epsilon_{\bf p}^3} \biggr) 
= {e^2 \over 4\pi}{m \over |\mu|}. 
\end{equation}

The absolute value of the Hall conductivity $\kappa$ 
shows Mt. Fuji shape\footnote{
Ishikawa tried to calculate $\kappa$ for non-zero $\mu$, 
but his result (3.3) in [I] is incorrect. 
We have corrected it in [NK]. 
Since we were interested in the case of $N_{\rm f}=2$ 
where $N_{\rm f}$ is the flavor degrees of freedom, 
we have obtained $2\kappa$ in [NK] and [NKF]. 
See, for example, the chapters 10 and 11 of \cite{F} 
on the chiral spin states and anyons with $N_{\rm f} \geq 2$. 
In the present Note I consider the case of $N_{\rm f}=1$. 
[I] $\equiv$ Ishikawa: 
Chapter-10 {\it Anomaly and Quantum Hall Effect} in 
{\it Quarks, Mesons and Nuclei: I. Strong Interactions} 
eds. Hwang and Henley (World Scientific, Singapore, 1989).  
[NK] $\equiv$ Narikiyo and Kuboki: J. Phys. Soc. Jpn. {\bf 62}, 1812 (1993).} 
as a function of the chemical potential $\mu$. 

\vskip 40pt
\noindent{\bf\Large A3. Ishikawa-Matsuyama formula}
\vskip 20pt

As discussed above the DC Hall conductivity is determined 
by the $q$-linear contribution of  
\begin{equation}
{\rm tr} \big[ \gamma_\mu G(p+q) \gamma_\nu G(p) \big]. 
\end{equation}
In the following we consider the case with zero chemical potential. 
Since 
\begin{equation}
{1 \over {\slashed p}+{\slashed q}-m} = 
{1 \over {\slashed p}-m} 
- {1 \over {\slashed p}-m} {\slashed q} {1 \over {\slashed p}-m} 
+ \cdot\cdot\cdot, 
\end{equation}
the $q$-linear contribution is 
\begin{equation}
- {\rm tr} \big[ 
\gamma_\mu G(p) \gamma_\lambda G(p) \gamma_\nu G(p) \big] q^\lambda. 
\label{3-gamma} 
\end{equation}
This is the result for free fermions. 

\vskip 4mm

\vskip 4mm
\begin{figure}[htbp]
\begin{center}
\includegraphics[width=12.6cm]{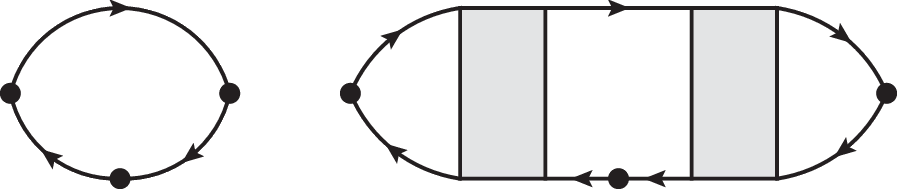}
\caption{
(Left): Insulator and (Right): Metal.} 
\label{fig:sigma_xy}
\end{center}
\end{figure}

In the case of interacting fermions 
the propagator $G(p)$ is renormalized into ${\tilde G}(p)$ 
and $\gamma$ into $\Gamma$ 
so that the DC Hall conductivity is determined by 
\begin{equation}
- {\rm tr} \big[ \Gamma_\mu {\tilde G}(p) 
\Gamma_\lambda {\tilde G}(p) \Gamma_\nu {\tilde G}(p) \big] q^\lambda, 
\label{Ren-Gamma} 
\end{equation}
whose diagrammatic representation is given as Fig.\ 4-(Left). 
The Ward identity\footnote{
See, for example, (19.20) of [BD].} 
tells us  
\begin{equation}
\Gamma_\mu = {\partial {\tilde G}(p)^{-1} \over \partial p^\mu}, 
\end{equation}
so that (\ref{Ren-Gamma}) becomes 
\begin{equation}
- {\rm tr} \biggl[ 
{\partial {\tilde G}(p)^{-1} \over \partial p^\mu} {\tilde G}(p)
{\partial {\tilde G}(p)^{-1} \over \partial p^\lambda} {\tilde G}(p)
{\partial {\tilde G}(p)^{-1} \over \partial p^\nu} {\tilde G}(p) \biggr] q^\lambda. 
\label{Gamma-G} 
\end{equation}
If the renormalized propagator\footnote{
See, for example, (19.18) of [BD]. 
Such a constant renormalization is relevant 
to the case of the DC conductivity 
which is governed by the lowest energy excitations.}
is given as 
$ {\tilde G}(p) = Z_2 G(p)$ 
with a constant $Z_2$, 
(\ref{Gamma-G}) becomes\footnote{
This form was introduced by Ishikawa and Matsuyama~\cite{IM} in 1980's 
and has been quoted in topological studies [H,V] in the 21st century. 
[H] $\equiv$ Haldane: Phys. Rev. Lett. {\bf 93}, 206602 (2004). 
[V] $\equiv$ Volovik: Physics Reports {\bf 351}, 195 (2001).} 
\begin{equation}
- {\rm tr} \biggl[ 
{\partial G(p)^{-1} \over \partial p^\mu} G(p)
{\partial G(p)^{-1} \over \partial p^\lambda} G(p)
{\partial G(p)^{-1} \over \partial p^\nu} G(p) \biggr] q^\lambda. 
\label{partial-G} 
\end{equation}
Since 
\begin{equation}
\gamma_\mu = {\partial G(p)^{-1} \over \partial p^\mu}, 
\end{equation}
(\ref{partial-G}) is equal to (\ref{3-gamma}). 
Namely, the DC Hall conductivity is not renormalized by the interaction.  

This absence of interaction-renormalization occurs 
for insulators where the energy spectrum has the mass gap. 
On the other hand, 
in the case of metals where the chemical potential is out of the mass gap 
and in the continuum, 
we have to consider the effect of damping 
whose diagrammatic representation\footnote{
This damping diagram is consistent 
with the Boltzmann equation and the Fermi liquid theory. 
The linear response theory of the DC Hall conductivity 
contains the other diagrams as discussed in [KY]. 
[KY] $\equiv$ Kohno and Yamada: Prog. Theor. Phys. {\bf 80}, 623 (1988).} 
is given as Fig.\ 4-(Right). 
The damping relevant to the DC conductivity 
is dominantly determined by the process 
with zero-energy excitation at the chemical potential. 
In the case of insulators 
such an excitation is absent so that the conductivity is not renormalized. 

\vskip 40pt
\noindent{\bf\Large A4. Remarks}
\vskip 20pt

\begin{enumerate}
\item 
The success of the Ishikawa-Matsuyama formula 
results from the cancelation between the self-energy and vertex corrections. 
On the other hand, 
the FLEX approximation violates such a cancelation 
as criticized in {\ttfamily arXiv:1406.5831}. 

\item
{\bf The derivation of the Chern-Simons action is 
nothing but the calculation of the anomalous Hall conductivity}. 
Here we give the discussion of the latter. 
Let us consider the Hamiltonian density 
\begin{equation}
H({\vec r}) = {\vec r} \cdot {\vec\sigma} .
\nonumber 
\end{equation}
If we take the parameter ${\vec r}$ as ${\vec r} = (p_1, p_2, m)$, 
this Hamiltonian leads to the Lagrangian (70). 
Now we take  ${\vec r} = (x, y, z)$ so that 
\begin{equation}
H({\vec r}) = x \sigma_x + y \sigma_y + z \sigma_z .
\nonumber 
\end{equation}
The energy eigenvalues of this Hamiltonian are $+r$ and $-r$ 
where $r = \sqrt{ x^2 + y^2 + z^2 }$. 
From the eigenvectors we can introduce the Berry connection $\vec A$. 
Also we can introduce the magnetic field ${\vec B} = \nabla \times {\vec A}$. 
The eigenvector for $\mp r$ leads to 
\begin{equation}
{\vec B} = \pm \frac{1}{2} \frac{\vec r}{r^3} .
\nonumber 
\end{equation}
This represents the magnetic field of the monopole with the charge $\pm 1/2$. 
In our case of ${\vec r} = (p_1, p_2, m)$ 
the anomalous Hall conductivity $\sigma_{\rm AHC}$ is given by (5.13) in [Vanderbilt] 
where $\sigma_{\rm AHC}$ is determined by the integral of $B_3$ over $(p_1, p_2)$ 
satisfying $\epsilon_{\bf p} < \mu$. 
When $\mu < -m$, the integrant $B_3$ for $-E_{\bf p}$ is positive 
so that $\sigma_{\rm AHC}$ is a increasing function of $\mu$. 
When $-m < \mu < m$, $\sigma_{\rm AHC}$ is saturated and becomes a constant. 
When $m < \mu$, the negative contribution $B_3$ for $E_{\bf p}$ is added to the constant 
so that $\sigma_{\rm AHC}$ is a decreasing function of $\mu$. 
The resulting $\sigma_{\rm AHC}$ is the same as our (88), (95) and (96). 
Such a result is reported in [Sinitsyn et al.] 
where {\bf the so-called Fermi-surface contribution is absent}. 
The reason for the absence is discussed in the following. 

\vskip 6pt
[Vanderbilt] {\it Berry Phases in Electronic Structure Theory}  (Cambridge Univ. Press, 2018) 

[Sinitsyn et al.] PRL {\bf 97}, 106804 (2006)  

\vskip 6pt

\item
Here we consider the non-relativistic linear response theory. 
The conductivity $\sigma_{xy}$ is given by (1) in [HT] in the absence of magnetic field. 
The so-called Fermi-surface contribution appears 
if we replace 
\begin{equation}
\frac{ n_{\rm F}\big( \epsilon_n({\bf k}) \big)  - n_{\rm F}\big( \epsilon_m({\bf k}) \big)  }
{ \epsilon_n({\bf k}) - \epsilon_m({\bf k}) }
\nonumber 
\end{equation}
by 
\begin{equation}
\frac{ {\rm d} n_{\rm F}\big( \epsilon_n({\bf k}) \big) }{ {\rm d} \epsilon_n({\bf k}) } 
\nonumber 
\end{equation}
for $n=m$ as discussed in [HT]. 
This contribution is relevant if we consider the Drude conductivity in the presence of dissipation 
as shown by (4) and (5) in [HT]. 
However, in our case of dissipationless anomalous Hall conductivity such a contribution is absent. 
In the next item the absence is discussed in the relativistic linear response theory. 

\vskip 6pt
[HT] Physical Review B {\bf 108}, 155108 (2023)

\vskip 6pt

\item
The relativistic linear response theory 
for the  conductivity $\sigma_{xy}$ in the absence of magnetic field is discussed in [Matsuyama]. 
The non-relativistic counterpart of the retarded polarization function (3.18) in [Matsuyama] 
is easily obtained by the use of  the spectral representation 
as shown in arXiv: 1112.1513v2 (See (172)). 
The conductivity (4.14) in [Matsuyama] is obtained for vanishing $\omega$ 
but we obtain the relativistic counterpart of  (1) in [HT] if we retain $\omega$. 
In (4.14) in [Matsuyama] the factor 
$ [n_{\rm F}(\omega_1) - n_{\rm F}(\omega_2)]/[\omega_1 - \omega_2]$ appears. 
However, the denominator of this factor is cancelled 
by the factor $(A^0-B^0)$ in the trace seen in the equation next to (4.14). 
Thus, there is no chance to introduce $ {\rm d}n_{\rm F}(\omega_2) / {\rm d}\omega_2 $ 
so that $ f(-\mu \pm C, -\mu \pm C) $ vanishes exactly. 
If we use  $ f(-\mu \pm C, -\mu \pm C) = 0 $, 
we obtain the correct result, (88), (95) and (96), from (4.14). 
The Fermi-surface contribution is absent, $ f(-\mu \pm C, -\mu \pm C) = 0 $, 
in the dissipationless anomalous Hall conductivity 
in consistent with the non-relativistic case in the previous item. 

\vskip 6pt
[Matsuyama] Bull. Nara Univ. Educ. {\bf 66}, 13 (2017)

\vskip 6pt

\item
In the previous item we have pointed out the incorrect calculation by [Matsuyama]. 
Here we point out another incorrect calculation by [SSS]. 
The analyticity of the propagator given by (1) in [SSS] agrees with ours (94). 
However, the polarization function is calculated by propagators 
with different incorrect analyticity seen in (4) of [SSS]. 
Thus, the obtained Chern-Simons term for $\mu^2 > m^2$ is incorrect 
both in [Matsuyama] and in [SSS]. 

\vskip 6pt
[SSS] arXiv:hep-th/9612140v2

\vskip 6pt

\item 
We have calculated $\kappa$ by the singularity above the real axis. 
The same result is also obtained by the singularity below the real axis. 
For example, the result of (96) for $\mu > m$ is easily obtained as 
\begin{equation}
\kappa = {me^2 \over 4\pi}
\int_{\mu}^{\infty} {\epsilon_{\bf p}d\epsilon_{\bf p} \over \epsilon_{\bf p}^3} 
= {e^2 \over 4\pi}{m \over |\mu|}. 
\nonumber
\end{equation}

\item
The result of (96) for $\mu > m$ is also obtained by 
\begin{equation}
\kappa = {me^2 \over 4\pi}
\int_{-\infty}^{-\mu} {-\epsilon_{\bf p}d\epsilon_{\bf p} \over (-\epsilon_{\bf p})^3} 
= {e^2 \over 4\pi}{m \over |\mu|}
\nonumber
\end{equation}
where we have taken into the fact 
that the excitation of a particle-hole pair is forbidden 
in the energy range between $-\mu$ and $\mu$. 

\item
The above $\kappa$ is expressed in the following compact form: 
\begin{equation}
\kappa = {e^2 \over 4\pi}{m \over {\rm max} \{ m, |\mu| \} }.
\nonumber
\end{equation}

\item
Semi-classically $\sigma_{\rm AHC}$ is given by (5.13) in [Vanderbilt] 
as mentioned in item 2. 
This formula results from the equation of motion (5.11a) 
with the anomalous velocity related to the Berry curvature 
and is the integral over the whole Fermi sea. 
\underline{ Surprisingly [SMJDS] } 
the formula can be transformed into the integral on the Fermi surface. 

\vskip 6pt
[SMJDS] Phys. Rev. B {\bf 75}, 045315 (2007)

\vskip 6pt

\item
The Kubo formula also gives the anomalous Hall conductivity 
represented by the integral over the whole Fermi sea as (A5) in [CB]. 
Using integration by parts the contribution of the Fermi sea 
can be replaced by that of the Fermi surface as discussed in Appendix A of [CB]. 

\vskip 6pt
[CB] Phys. Rev. B {\bf 64}, 014416 (2001)

\vskip 6pt

\item
Karplus and Luttinger [KL] already noticed 
that the contributions from the Fermi sea and the Fermi surface are interchangeable\footnote{
See item 9 and item 10.
}. 

\vskip 6pt
[KL] Phys. Rev. {\bf 95}, 1154 (1954)

\vskip 6pt

\item
Related to item 3, the polarization function (See (30.9) in [FW]) 
in the non-relativistic linear response theory is given by 
\begin{equation}
\sum_{\bf k}
\frac{ n_{\rm F}\big( \epsilon({\bf k}+{\bf q}) \big)  - n_{\rm F}\big( \epsilon({\bf k}) \big)  }
{ \epsilon({\bf k}+{\bf q}) - \epsilon({\bf k}) - (\omega +  i \delta) }.
\nonumber 
\end{equation}
In the calculation of the anomalous Hall conductivity 
we need the polarization function for ${\bf q}={\bf 0}$ and $\omega \neq 0$ 
so that we do not encounter the derivative of $n_{\rm F}$. 
If we consider another case where $\omega = 0$ and ${\bf q} \rightarrow {\bf 0}$, 
we encounter the derivative of $n_{\rm F}$ 
which represents the contribution of the Fermi surface. 
However, it differs from the derivative of $n_{\rm F}$ obtained by integration by parts 
discussed in item 10. 

\vskip 6pt
[FW] {\it Quantum Theory of Many-Particle Systems}  (McGraw-Hill, 1971) 

\vskip 6pt

\item
In the above item 2 we consider the Berry phase in the absence of the electromagnetic field\footnote{
We have calculated the polarization function (81) in the absence of the electromagnetic field. 
Thus the obtained $\gamma_\mu \gamma_\lambda \gamma_\nu$ term 
represents the {\bf intrinsic} polarization of the whole Dirac sea [FS]. 
} where the parameter in the Hamiltonian is the momentum. 
Since the energy eigenvalue is determined by $\Pi^k = p^k + e A^k$, 
the Berry phase in the presence of the electromagnetic field is obtained 
if we regard the vector potential as the parameter 
where the vector potential plays the same role 
as that of the momentum in the absence of the electromagnetic field [FS]. 
Consequently the Chern-Simons effective action is given by the Berry phase 
integrated over the whole Dirac sea [FS]. 

\vskip 6pt
[FS] Phys. Rev. B {\bf 49}, 4277 (1994)

\end{enumerate}



\begin{thebibliography}{99}
\bibitem{xxx} 
In the following 
I only list the references that you must read. 
Neither originality nor priority is considered here. 
Other references are cited in the footnotes.
\end{thebibliography}
\end{document}